\documentclass{article}

\usepackage{authblk}

\usepackage{graphicx} 
\usepackage{amsmath,amsthm,amssymb,amsfonts,latexsym}
\usepackage{url}
\usepackage{comment}
\usepackage{float} 
\usepackage[left=2.5cm,top=2.5cm,right=2.5cm,bottom=2cm]{geometry}

\usepackage{color}
\usepackage{subcaption}

\usepackage{hyperref}

\newtheorem{theorem}{Theorem}[section]

\newtheorem{lemma}[theorem]{Lemma}
\newtheorem{proposition}[theorem]{Proposition}
\newtheorem*{remark}{Remark}

\newcommand{\Q}{\mathbb{Q}}
\newcommand{\R}{\mathbb{R}}
\newcommand{\N}{\mathbb{N}}
\newcommand{\E}{\mathbb{E}}
\newcommand{\D}{\mathrm{d}}

\usepackage{titlesec}
\titleformat{\paragraph}
  {\normalfont\normalsize\bfseries}{\theparagraph}{1em}{}
\titlespacing*{\paragraph}{0pt}{2ex}{1ex}

\title{Quantum computing for multidimensional option pricing: End-to-end pipeline}

\author[1]{Julien Hok}
\author[2,3,*]{Álvaro Leitao}

\affil[1]{Investec Bank, UK}
\affil[2]{CITIC Research center, Spain}
\affil[3]{Department of Mathematics, University of A Coruña, Spain}
\affil[*]{Corresponding author: alvaro.leitao@udc.gal}
\date{January 8, 2026}

\begin{document}

\maketitle

\begin{abstract}
This work introduces an end-to-end framework for multi-asset option pricing that combines market-consistent risk-neutral density recovery with quantum-accelerated numerical integration. We first calibrate arbitrage-free marginal distributions from European option quotes using the Normal Inverse Gaussian (NIG) model, leveraging its analytical tractability and ability to capture skewness and fat tails. Marginals are coupled via a Gaussian copula to construct joint distributions. To address the computational bottleneck of the high-dimensional integration required to solve the option pricing formula, we employ Quantum Accelerated Monte Carlo (QAMC) techniques based on Quantum Amplitude Estimation (QAE), achieving quadratic convergence improvements over classical Monte Carlo (CMC) methods. Theoretical results establish accuracy bounds and query complexity for both marginal density estimation (via cosine-series expansions) and multidimensional pricing. Empirical tests on liquid equity entities (Credit Agricole, AXA, Michelin) confirm high calibration accuracy and demonstrate that QAMC requires 10–100 times fewer queries than classical methods for comparable precision. This study provides a practical route to integrate arbitrage-aware modelling with quantum computing, highlighting implications for scalability and future extensions to complex derivatives.
\end{abstract}

\section{Introduction}

Pricing options on multiple underlying assets is a central problem in quantitative finance, with broad relevance for risk management, structured products, and trading of multi-asset exotics. In high dimensions, classical valuation workflows (spanning construction of risk-neutral distributions, consistent interpolation/extrapolation of market surfaces, and numerical integration of complex payoffs) face significant computational and modelling challenges. A central requirement is the recovery of arbitrage-free marginal risk-neutral densities from observed vanilla options and their implied volatilities, together with a tractable and realistic representation of inter-asset dependence to obtain joint distributions suitable for pricing basket, spread, worst-of, and other path-independent multivariate payoffs. In this context, traditional approaches rely heavily on simplistic stochastic models and numerical techniques such as classical Monte Carlo (CMC) simulation, which, while robust, often suffer from lack of representativeness and high computational costs when extended to high-dimensional settings. 

Two strands of progress have shaped this landscape. First, on the distributional modelling side, the industry has moved beyond lognormal assumptions, motivated by empirical features such as negative skew and fat tails in equity returns. L\'evy models capture jumps and heavy tails while retaining analytical tractability through characteristic functions, facilitating Fourier-based valuation~\cite{CarrMadan1999, Schoutens2003, FangOosterlee2008, HokChan2017}. Within this class, the Normal Inverse Gaussian (NIG) model is particularly attractive: smooth densities, tuneable skew/kurtosis, arbitrage-free time-slice calibration, and empirically validated fits to equity options~\cite{ghysels2009nig,maekawa2004option}. These properties make NIG well-suited for constructing marginal market distributions required for multi-asset pricing. In contrast, the CMC simulation of the NIG process is not trivial and rather inefficient, specially in high dimensions. Regarding the dependence structure, in finance, it is often modelled separately from the marginals by employing copulas \cite{Cherubini2004}. The well-known Sklar's theorem guarantees that a copula combined with marginals yields a valid joint distribution~\cite{nelsen2006copulas}. Second, On the computational side, CMC methods have long been the workhorse for option pricing (see \cite{Glasserman2004}), but their slow convergence rate, $O(1/\epsilon^2)$), poses challenges for high accuracy in large dimensions. Native quantum algorithms, particularly those proposed in \cite{Brassard2002,Montanaro2015,Rebentrost_2018}, exploit the so-called Quantum Amplitude Estimation (QAE) to achieve $O(1/\epsilon)$ convergence, offering a theoretical quadratic improvement. The confluence of these strands raises a compelling question: can a market-data-driven, arbitrage-aware construction of multi-asset pricing distributions be paired with quantum-accelerated estimators to achieve practical gains in accuracy-vs-cost for multidimensional option pricing? Before describing our proposal to address this question, let us discuss some related literature review.

In order to encapsulate the market information of each individual asset, practitioners typically work with implied volatility surfaces, motivating robust interpolation/extrapolation that avoids static arbitrage. The Stochastic Volatility Inspired (SVI) parameterization~\cite{gatheral2004svi} and its arbitrage-free extensions~\cite{gatheral2014svi} are widely adopted due to parsimony and control over convexity and butterfly arbitrage. Alternatives include local volatility bootstrapping and tied time-dependent parameters~\cite{andreasen2011interpolation,lipton2011filling}, stochastic volatility families such as ~\cite{Heston1993, hagan2002sabr}, and all-maturities non-parametric approaches imposing global no-arbitrage constraints~\cite{demarch2019sinkhorn}. Regularization techniques (e.g., Tikhonov) are standard for stabilizing ill-posed calibration~\cite{ContTankov2004,Crepey2003}. The calibration of some of the aforementioned models is treated in, for example, \cite{FernandezEtAl2013, HokTan2019,Leitao2021}.

Quantum computing explores how the principles of quantum mechanics can be harnessed to enhance information processing beyond classical limits. Since its inception, the field has witnessed remarkable progress in algorithm design and hardware development, driving rapid growth in quantum technologies and fuelling the search for practical applications across diverse domains. Among these emerging areas, quantitative finance has attracted significant attention as a promising candidate for quantum-enabled innovation, see \cite{Orus2019, Gomez_2022} and the references therein. For the particular task of options pricing via Monte Carlo-like methods, recent works (see \cite{Stamatopoulos2020, Vazquez2020, mrqae, alonso2025}) have demonstrated practical pipelines for quantum-based approaches, including state preparation and encoding strategies. Within this framework, quantum advantage arises from applying the QAE routine to integral-based formulations, such as those used in option pricing. However, the original QAE implementation remains impractical under current hardware constraints. To address this limitation, several hardware-efficient variants have emerged in recent years, \cite{Grinko_2021, Fukuzawa_2023, rqae, mrqae} among others, enabling the deployment of QAE on near-term quantum devices. Still, most quantum computing demonstrations applied to financial derivatives problems use stylized distributions or toy payoffs. There is a lack of end-to-end pipelines that: (a) infer arbitrage-free risk-neutral marginals from real option quotes, (b) assemble joint distributions with empirically meaningful dependence, and (c) perform quantum-accelerated valuation.

Then, this paper addresses the previous points (so it tries to answer the question above) by presenting a full pipeline: (i) recovery of market-consistent marginal risk-neutral densities using the exponential NIG model, (ii) assembly of joint distributions via copulas (Gaussian copula for tractability), and (iii) deployment of a Quantum Accelerated Monte Carlo (QAMC) approach which acts on both the marginal density estimation (via orthogonal cosine expansions) and on the final multidimensional option valuation. The pipeline is modular (amenable to alternative marginals and copulas) and quantifies the accuracy–cost trade‑offs under both classical and quantum estimators. From the market distribution construction viewpoint, we provide relevant practical results (independence of prices from NIG location parameter, continuity and existence of regularized calibration solutions), arbitrage sanity checks, and empirical validation on liquid single-name equities (Credit Agricole, AXA, Michelin), which allows us to come up with calibrated distributions that match market skew and tails. In regard with the proposed quantum computing-based solution for multivariate pricing, we demonstrate, both theoretically and empirically, that QAMC achieves the expected quadratic convergence improvement compared with CMC when applied to crucial points in the whole pipeline, namely, the marginal distribution reconstruction and final multi-asset option valuation. In this sense, the choice of the NIG model is not arbitrary since, under its formulation, the density function driving the asset evolution present an analytical expression while the distribution and quantile functions (required for the CMC simulation) are not available in closed-form, resulting in computational expensive sampling procedures.

The paper is organized as follows. Section \ref{sec:market_distributions} describes how to construct the asset distributions from the market information, including details of procedural issues (Sections \ref{sec:market_call_puts} and \ref{sec:fitting_methods}), the exponential NIG model (Section \ref{sec:exponential_NIG_model}) and calibration results (\ref{sec:calibration_results}). In Section \ref{sec:option_pricing_quantum}, the different components of the quantum-based multidimensional option valuation are presented: the general pricing formula and the inclusion of copulas in it (Section \ref{sec:multidim_option_pricing_copulas}), the QAMC method applied to both marginals recovery and final option price calculation along with a rigorous theoretical analysis (Section \ref{sec:quantum_multidim}) and the experimental outcomes (Section \ref{sec:quantum_experimental}). Finally, Section \ref{sec:conclusions} concludes with a discussion of the main findings.

\section{Construction of the market distributions}\label{sec:market_distributions}

To construct a multidimensional market risk-neutral distribution using the copula framework, we begin by modelling the marginal distributions of each underlying asset. This step involves fitting a parametric distribution to European option prices observed in the market for each maturity. In this work, we adopt the NIG distribution introduced in Section~\ref{sec:exponential_NIG_model}.

\subsection{Market European call and put options prices}\label{sec:market_call_puts}

The prices of European call and put options under the risk-neutral measure, expressed in terms of the risk-neutral density \( f(S_T) \), are given by

\begin{align*}
C(T, K) &= e^{-rT} \int_K^{\infty} f(S_T) (S_T - K) \, \D S_T, \\
P(T, K) &= e^{-rT} \int_0^{K} f(S_T) (K - S_T) \, \D S_T,
\end{align*}
where $T$ option maturity, $K$ strike price, $r$ risk-free interest rate and $f(S_T)$ risk-neutral density of the underlying asset at maturity.

From Breeden and Litzenberger \cite{breeden1978prices} and assuming enough regularity, differentiating once with respect to \( K \) yields the cumulative distribution function,

\begin{equation*}
\frac{\partial C(T, K)}{\partial K} = -e^{-rT} \int_K^{\infty} f(S_T) \, \D S_T,
\end{equation*}
while differentiating twice produces the probability density function,

\begin{equation} \label{eq:MarketPDF}
\frac{\partial^2 C(T, K)}{\partial K^2} = e^{-rT} f(K).
\end{equation}

Usually, market vanilla prices are first converted to implied volatilities using the Black-Scholes options pricing formula,

\begin{equation*}
\begin{aligned}
C^{\text{BS}}(T,K) &= S_0 e^{-qT} \Phi(d_1) - K e^{-rT} \Phi(d_2), \\
P^{\text{BS}}(T,K) &= K e^{-rT} \Phi(-d_2) - S_0 e^{-qT} \Phi(-d_1),
\end{aligned}
\end{equation*}
where
\begin{equation*}
d_1 = \frac{\log\left(\frac{S_0}{K}\right) + \left(r - q + \frac{1}{2} \sigma^2 \right)T}{\sigma \sqrt{T}}, \quad d_2 = d_1 - \sigma \sqrt{T},
\end{equation*}
with $S_0$ the underlying asset spot price, $q$ the continuous dividend yield, $\sigma$ the volatility of the underlying asset and $\Phi(.)$ the cumulative distribution  function of the standard normal distribution. The implied volatility $\sigma_{\text{imp}}(T,K)$ associated to an expiry $T$ and strike $K$ is defined by matching market, $\bar{V}$, and Black-scholes prices, i.e,
\begin{equation*}
V^{\text{BS}}(T,K; \sigma_{\text{imp}}) = \bar{V}(T,K), \quad V \in \{C, P\},
\end{equation*}
which is well-defined by the strictly increasing Black-Scholes price with respect to the volatility parameter. So there is a one to one mapping between vanilla prices and the implied volatilities.  Practitioners represent vanilla options market data as implied volatility because it is easier to interpret and to monitor. For a given maturity, implied volatility as a function of strike is not constant and often smile shaped or skewed. It is usually called \textit{volatility smile}.

 To work with the risk-neutral density formula (\ref{eq:MarketPDF}) one needs the vanilla prices or implied volatilities at any
positive strike. Market data for traded options is only available at discrete strike points. As a consequence, we need an interpolation/extrapolation engine to produce a smooth function $\bar{C}(T,K) / \bar{P}(T,K)$ or $\sigma_{\text{imp}}(T,K)$, given a discrete market data set, which is discussed in the next section. 

\subsection{Fitting and interpolation methods}\label{sec:fitting_methods}

The available fitting methods for implied volatility surfaces can be grouped into several categories, depending on how the market information is represented. These include:

\begin{itemize}

    \item Implied-volatility -- describes directly the implied volatility as a function of strike and maturity. Different parametrizations are used across the industry, from simplistic (and arbitrageable) quadratic skew with cutoffs, to splines or to SVI parametrization \cite{gatheral2004svi, gatheral2014svi}.

    \item Time-slice distribution -- defines the distribution of the stock price independently for every maturity. Typical examples include usage of a stochastic volatility model generated distribution, like the SABR model \cite{hagan2002sabr}, or directly a parametrization of the stock probability density function.

    \item Non-homogeneous stochastic process -- bootstraps time-dependent parameters of a stochastic process by fitting the implied volatilities at each maturity chronologically. A good example is the tied local volatility approach introduced in \cite{andreasen2011interpolation}  and improved in \cite{lipton2011filling}.

    \item All-maturities non-parametric density -- fits all the maturities together in a non-arbitrageable way. An interesting approach has been developed in \cite{demarch2019sinkhorn}.

\end{itemize}
All these methods have their advantages and disadvantages that have been discussed in e.g. \cite{demarch2019sinkhorn}. In this work, we adopt the time-slice distribution method, using the NIG distribution to parametrize the risk-neutral probability density function of the underlying asset at each maturity as illustration. Others methods discussed above can also be used to build the market distributions.  

\subsection{The exponential NIG model} \label{sec:exponential_NIG_model}

Let $T > 0$ be a fixed time horizon, and let $S : t \in [0, T] \mapsto S(t)$ denote the market price of a financial asset. We assume that, under the risk neutral probability $\Q$, the dynamics of $S(t)$ follow
\begin{equation}
\frac{\D S(t)}{S(t)} = (r - q) \, \D t + \D X(t), \quad t \in [0, T], \label{eq:expNIG_SDE}
\end{equation}
where the initial value is given, i.e., $S(0) = S_0$, $r \geq 0$ is the risk-free interest rate, $q \geq 0$ is the continuous dividend yield (both deterministic and continuously compounded) and $\{X(t)\}_{t \in [0,T]}$ is a NIG Levy process with $X(0)=0$, whose increments satisfy
\begin{equation*}
    X(t+\Delta t) - X(t) \sim \text{NIG}(\alpha, \beta, \delta \Delta t, \mu \Delta t) \quad \text{for all } \Delta t \geq 0,
\end{equation*}
with the NIG distribution with parameters  $(\alpha, \beta, \delta, \mu)$, written as $\text{NIG}(\alpha, \beta, \delta, \mu)$, has the following density function,
\begin{equation}\label{eq:NIG_density}
f_{\mathrm{NIG}}(x; \alpha, \beta, \delta, \mu) = \frac{\alpha\delta}{\pi} e^{\delta\sqrt{\alpha^2 - \beta^2} + \beta(x - \mu)} \frac{\mathrm{K}_1\left(\alpha\sqrt{\delta^2 + (x - \mu)^2}\right)}{\sqrt{\delta^2 + (x - \mu)^2}}, \hspace{0.5cm} x \in \R
\end{equation}
where:
\begin{itemize}
    \item $\mathrm{K}_1(z)$ is the modified Bessel function of the second kind with index 1,
    \item $\alpha > 0$ is the tail (steepness) parameter which controls the kurtosis (larger $\alpha$ gives lighter tails),
    \item $\beta \in (-\alpha, \alpha)$ is the skewness parameter ($\beta < 0$ implies left skewness, $\beta > 0$ right skewness, and $\beta = 0$ yields symmetry),
    \item $\delta > 0$ is the scale parameter,
    \item $\mu \in \mathbb{R}$ is the location parameter.
\end{itemize}

The NIG process characteristic function $\varphi(u; t) := \mathbb{E}\left[e^{iu X_t}\right]$ can be written down as 
$\varphi(u; t) = e^{t \cdot \vartheta(u)}$,
where the characteristic exponent or Lévy symbol, is known in exact form

\begin{equation*}
\vartheta(u) := \log \varphi(u; 1) = i \mu u - \delta \left( \sqrt{\alpha^2 - (\beta + iu)^2} - \sqrt{\alpha^2 - \beta^2} \right).
\end{equation*}

The solution to the stochastic differential equation \eqref{eq:expNIG_SDE} is given by the exponential NIG process

\begin{equation*}
S(T) = S(t) \exp\left( (r - q + \omega)\tau + X(\tau) \right), \quad \tau := T - t, \label{eq:expNIG_solution}
\end{equation*}
where $\omega$ is the \textit{martingale or compensator adjustment}. It ensures that the discounted asset price is a true $\mathbb{Q}$-martingale, by enforcing
\[
\mathbb{E}^{\mathbb{Q}}[S(T) \mid \mathcal{F}_t] = e^{(r - q)\tau} S(t).
\]
which leads to the condition
\begin{equation*}
\omega = -\mu + \delta \left( \sqrt{\alpha^2 - (\beta + 1)^2}  - \sqrt{\alpha^2 - \beta^2} \right). \label{eq:omega_exact}
\end{equation*}

The NIG distribution enjoys the following desirable properties in our context:

\begin{itemize}
    \item It admits an explicit density function, which is smooth and differentiable, ensuring numerical stability.
    
    \item It is arbitrage-free across time slices when calibrated individually per maturity.
    
    \item The characteristic function is known in closed form, enabling efficient pricing via Fourier inversion techniques.
    
    \item Its flexible tail behaviour and skewness allow it to fit market-implied distributions accurately (see e.g \cite{ghysels2009nig, tavin2012implied, maekawa2004option}).
\end{itemize}

By obtained a set of calibrated NIG parameters $(\bar{\alpha}, \bar{\beta}, \bar{\delta}, \bar{\mu})$ for each maturity $T$ to the observed market option prices or implied volatilities, we can recover a smooth, arbitrage-free risk-neutral density $\bar{f}_{\mathrm{NIG}}(x)$. This calibrated NIG density serves as the marginal distribution for the asset price at maturity $T$, and will be later coupled across assets using a copula function as described in Section \ref{sec:copula}.

\subsubsection{NIG model calibration} \label{sec:NIG_calibration}

In the following, some useful results on the calibration of the NIG model are provided.

\begin{proposition}[Independence of the NIG price on the location parameter]
\label{prop:independence_location}
Given the NIG pricing model in Section \ref{sec:exponential_NIG_model} and let $h: \mathbb{R}_{+} \to \mathbb{R}$ be a measurable payoff function (e.g., a European call or put payoff) such that the European option price,
$$
V^{\mathrm{NIG}}(T, K; \theta) = e^{-rT} \mathbb{E}^{\mathbb{Q}}[h(S(T), K)],
$$
is well-defined and finite. Then $V^{\mathrm{NIG}}(T, K; \theta)$ is independent of the location parameter $\mu$.
\end{proposition}

\begin{proof}
The log-price at time $T$ can be expressed as
$$ \log S(T) = \log S_0 + (r - q + \omega) T + X(T), $$
where $X(T) \sim \mathrm{NIG}(\alpha, \beta, \delta T, \mu T)$.

The density of $X(T)$ depends on $\mu T$ as a location shift.
The martingale correction $\omega$ is explicitly given by
$$ \omega = -\mu + \delta \left( \sqrt{\alpha^2 - (\beta + 1)^2} - \sqrt{\alpha^2 - \beta^2} \right), $$
which depends linearly on $-\mu$.

Substituting, the random variable $\log S(T)$ can be rewritten as
\begin{align*}
\log S(T) &= \log S_0 + (r - q) T + \omega T + X(T) \\
&= \log S_0 + (r - q) T + T \left( -\mu + \delta \left( \sqrt{\alpha^2 - (\beta + 1)^2} - \sqrt{\alpha^2 - \beta^2} \right) \right) + X(T) \\
&= \log S_0 + (r - q) T + \delta T \left( \sqrt{\alpha^2 - (\beta + 1)^2} - \sqrt{\alpha^2 - \beta^2} \right) + (X(T) - \mu T).
\end{align*}
Since $X(T) - \mu T \sim \mathrm{NIG}(\alpha, \beta, \delta T, 0)$, the distribution of $\log S(T)$ under $\mathbb{Q}$ depends only on $\alpha, \beta, \delta$ and not on $\mu$. Hence, the distribution of $S(T)$ and therefore the expectation $\mathbb{E}^{\mathbb{Q}}[h(S(T), K)]$ are independent of $\mu$.

\end{proof}

This result justifies fixing $ \mu = 0$ during calibration without loss of generality. To prepare for the proof of Proposition~\ref{prop:ExistenceMinSol}, we first establish that the model option prices are continuous with respect to the NIG parameters.

\begin{lemma}[Continuity of NIG Option Prices] \label{lm:ContinuityNIGEuroCallPutPrice}
	\noindent Let $V_m^{\mathrm{NIG}}(T, K; \theta)$ denote the price of the $m$-th call/put European option under the NIG model with parameter $\theta = (\alpha, \beta, \delta) \in \Theta$, with $\Theta$ a non empty compact set defined in Proposition \ref{prop:ExistenceMinSol}
Then, for each $m$, the mapping
$$
\theta \mapsto V_m^{\mathrm{NIG}}(T, K; \theta)
$$
is continuous on $\Theta$.
\end{lemma}

\begin{proof}
Let $\theta=(\alpha,\beta,\delta)\in\Theta$, and consider the price of the $m$-th European option under the NIG model,
$$
V_m^{\mathrm{NIG}}(T, K; \theta) = e^{-rT} \int_{\mathbb{R}} h\left(S_0 e^{\omega(\theta) T + x}, K\right) f_{\mathrm{NIG}}(x; \theta)\, \D x, 
$$
where $f_{\mathrm{NIG}}(x; \theta)$ is the NIG density with parameters $(\alpha, \beta, \delta T, 0)$, and $\omega(\theta)$ is the martingale correction term.
The map $\theta \mapsto \omega(\theta)$ is continuous, and $f_{\mathrm{NIG}}(x; \theta)$ is jointly continuous in $(x, \theta)$ on $\mathbb{R} \times \Theta$. Hence, the integrand is pointwise continuous in $\theta$ for each fixed $x \in \mathbb{R}$.
To apply the Dominated Convergence Theorem, we note that the admissibility conditions $\alpha^2 > \beta^2$ and $\alpha^2 > (\beta + 1)^2$ ensure that the NIG density decays exponentially in $|x|$, uniformly over $\theta \in \Theta$. For European call options, the payoff behaves like $h(S, K) \sim S \sim e^x$, so the integrand satisfies
$$
h(S_0 e^{\omega T + x}) \cdot f_{\mathrm{NIG}}(x; \theta) \sim e^{(1 + \beta)x - \alpha |x|}, 
$$
which is integrable when $\alpha > \beta + 1$. For put options, the payoff is bounded, and integrability follows directly from the exponential decay of $f_{\mathrm{NIG}}$.
Therefore, the integrands are uniformly dominated by an integrable function independent of $\theta$, and the Dominated Convergence Theorem yields
$$
\lim_{s \to \infty} V_m^{\mathrm{NIG}}(T, K; \theta_s) = V_m^{\mathrm{NIG}}(T, K; \theta),
$$
for any sequence $\theta_s \to \theta$ in $\Theta$. This proves continuity of $\theta \mapsto V_m^{\mathrm{NIG}}(T, K; \theta)$.
\end{proof}

\begin{proposition}[Existence of Solution to the Regularized Calibration Problem] \label{prop:ExistenceMinSol}
Let $\Theta \subset \mathbb{R}^3$ be a non-empty, compact subset of admissible parameters $\theta := (\alpha, \beta, \delta)$ for the NIG model with fixed $\mu := 0$, satisfying the constraints
\[
\alpha > 0, \quad \delta > 0, \quad \beta^2 < \alpha^2, \quad (\beta + 1)^2 < \alpha^2.
\]
Define the Tikhonov-regularized least-squares objective function,
\begin{equation*}
\mathcal{J}(\theta) := \sum_{m=1}^M w_m \left( V_m^{\mathrm{NIG}}(T, K; \theta) - \bar{V}_m(T, K) \right)^2 + \lambda \| \theta - \theta_0 \|^2,
\end{equation*}
where $\{ \bar{V}_m(T, K) \}_{m=1}^M$ are observed market European call/put option prices, $V_m^{\mathrm{NIG}}(T, K; \theta)$ are model European call/put option prices under the NIG model with parameter $\theta$, $w_m \geq 0$ are fixed weights, $\theta_0 \in \Theta$ is a fixed prior (reference) parameter vector, $\lambda \geq 0$ is the regularization parameter and $\| \cdot \|$ is the Euclidean norm on $\mathbb{R}^3$.

Then the minimization problem
\[
\min_{\theta \in \Theta} \mathcal{J}(\theta)
\]
admits at least one solution.
\end{proposition}

\begin{proof}
The parameter set $\Theta$ is compact, and all quantities in the objective function are finite by assumption. By Lemma~\ref{lm:ContinuityNIGEuroCallPutPrice}, the map $\theta \mapsto V_m(T, K; \theta)$ is continuous for each $m$. Therefore, $\mathcal{J}(\theta)$ is a continuous real-valued function on a compact domain. By the Weierstrass Extreme Value Theorem, $\mathcal{J}$ attains a global minimum on $\Theta$.
\end{proof}

\begin{remark}[Non-uniqueness]
The existence of a solution to the regularized calibration problem does not imply uniqueness. The objective function $\mathcal{J}(\theta)$ is generally non-convex due to the nonlinear dependence of option prices on the NIG parameters. Multiple local minima may exist, and standard optimization algorithms may converge to different solutions depending on the initial guess. 

\end{remark}

\begin{remark}[Stability and sensitivity]
While Proposition \ref{prop:ExistenceMinSol} guarantees the existence of a minimizer, the stability of the solution with respect to perturbations in the market data $\{ \bar{V}_m \}$ is not addressed. In ill-posed inverse problems such as model calibration, small changes in the input can result in large variations in the estimated parameters. The Tikhonov regularization term $\lambda \| \theta - \theta_0 \|^2$ is introduced precisely to mitigate such instability by enforcing proximity to a reference parameter $\theta_0$. The choice of $\lambda > 0$ thus balances calibration accuracy and stability (see e.g Chapter 3, Section 13 in \cite{ContTankov2004}, \cite{ContTankovVoltchkova2004} or \cite{Crepey2003} for more details).
\end{remark}

\subsection{Calibration methodology and practical implementation}\label{sec:calibration_results}

\subsubsection{Market data}\label{sec:market_data}

The market data used in our numerical experiments consists of European call and put option quotes on Credit Agricole, AXA, and Michelin (three major French companies) sourced from Euronext as of 24/12/2024. The dataset spans multiple maturities and, for each expiry, includes strike levels and bid-ask quotes for European call and put options. In addition, the data provides stock futures curves. Spot prices are taken from market closing levels retrieved via Yahoo Finance.

\subsubsection{Market-implied discount, forward, and dividend curves}

In our methodology, we construct market-implied discount factors, forward prices, and dividend yields to ensure consistency with observed European vanilla option prices. This step is essential to transition from market quotes to risk-neutral distributions used in option pricing and density recovery.

We leverage the classical \emph{put-call parity} relation for European vanilla options. For a given strike \( K \) and maturity \( T \), the parity reads
\begin{equation}\label{eq:putcallparity}
C(T, K) - P(T, K) = FW(T) \cdot DF(T) - K \cdot DF(T),
\end{equation}
where \( C(T, K) \) and \( P(T, K) \) denote the market prices of the European call and put options, respectively, \( FW(T) \) is the forward price of the underlying asset at maturity \( T \) and \( DF(T) \) is the risk-free discount factor at maturity \( T \).

From the bid and ask quotes, we compute mid-prices for European call and put options. We then apply equation~\eqref{eq:putcallparity} to perform a linear regression in the strike \( K \). The slope and intercept of this regression allow us to estimate the discount factor \( DF(T) \) and the forward price \( FW(T) \) for each expiry.

Using the inferred forward price, we deduce the continuous dividend yield \( q \) using the standard spot-forward relationship
\begin{equation*}
FW(T) = S_0 e^{(r - q)T},
\end{equation*}
where \( S_0 \) is the spot price and \( r \) is the risk-free interest rate.

This procedure ensures internal consistency across the inferred market curves and aligns all inputs (spot prices, forwards, and discounting factors) with actual observed option market data. By doing so, we avoid relying on external estimates of interest rates or dividend yields, which could introduce inconsistencies or arbitrage opportunities.

\subsubsection{Arbitrage sanity check}

Our pricing model is grounded in arbitrage-free principles. Accordingly, it is crucial that the input data exhibit internal consistency. For each option expiry, we verify the absence of digital and butterfly arbitrage, removing any violations prior to model calibration.


For a strictly increasing sequence of strikes \( K_1 < K_2 < K_3 \), we approximate digital call and put prices using one-sided finite differences of vanilla option prices. The absence of arbitrage requires that the following inequalities hold,
\begin{align*}
0 &< \frac{C(T, K_1) - C(T, K_2)}{K_2 - K_1} < 1,  \\
0 &< \frac{P(T, K_2) - P(T, K_1)}{K_2 - K_1} < 1.
\end{align*}

These expressions correspond to the implied prices of digital calls and puts, which must lie strictly between 0 and 1 under the no-arbitrage assumption.


Butterfly arbitrage arises when the option price surface fails to exhibit convexity in strike. For European call options, the following convexity condition must be satisfied,
\begin{equation*}
C(T, K_1) - C(T, K_2) - \frac{K_2 - K_1}{K_3 - K_2} \left(C(T, K_2) - C(T, K_3)\right) \geq 0.
\end{equation*}

An analogous condition applies to European put options, i.e.,
\begin{equation*}
P(T, K_1) - P(T, K_2) - \frac{K_2 - K_1}{K_3 - K_2} \left(P(T, K_2) - P(T, K_3)\right) \geq 0.
\end{equation*}

Violations of these conditions imply inconsistency in the implied risk-neutral probability distribution. In our dataset, we detect a small number of violations, specifically, digital put arbitrage in the far tails. These inconsistencies have been removed prior to the fitting procedure.

\subsubsection{Calibration methodology}

We calibrate the NIG distribution to market option prices for each asset and maturity by solving a regularized, constrained nonlinear least squares problem as presented in Proposition \ref{prop:ExistenceMinSol}. Recall that, this procedure is designed to recover smooth and arbitrage-free marginal risk-neutral densities, suitable for copula-based joint distribution construction.


Next, we adapt the result from Proposition \ref{prop:ExistenceMinSol} accounting for the methodological considerations described above. Let $\theta = (\alpha, \beta, \delta)$ denote the NIG parameters to be calibrated, with location $\mu$ fixed to zero (justified analytically in Proposition \ref{prop:independence_location}). Then, the calibration minimizes the following objective function,
\begin{equation*}
\mathcal{J}(\theta) = \sum_{m=1}^{M} w_m \left(V_m^{\text{NIG}}(T, K; \theta) - \bar{V}_m(T, K)\right)^2 + \lambda \|\theta - \theta_0\|^2,
\end{equation*}
where:
\begin{itemize}
  \item $\bar{V}_m(T, K)$ are the mid-market prices of liquid European call/put options with various strikes and maturities,
  \item $V_m^{\text{NIG}}(T, K; \theta)$ are again model prices computed using the exponential NIG model,
  \item $w_m$ are weights inversely proportional to the bid-ask spreads (to reflect pricing uncertainty),
  \item $\theta_0$ is a prior guess for the parameters and chosen to be given by the Black-Scholes model using the ATM implied volatility, and
  \item $\lambda \geq 0$ is a regularization coefficient that controls proximity to the prior.
\end{itemize}

This objective balances accuracy to market data with stability, following Tikhonov-style regularization as discussed in Cont and Tankov~\cite{ContTankov2004}.
Since the objective function is non-convex, careful initialization is essential. We perform a grid search over plausible starting points for $(\alpha, \beta, \delta)$, selecting the one with the lowest pre-optimization objective value. This heuristic helps mitigate convergence to poor local minima.
As optimizer, we use the \texttt{trust-constr} algorithm from \texttt{scipy.optimize.minimize}, which supports constraints, bounds, and robust convergence settings. Optimization tolerances are tightened to ensure precise convergence. This yields a calibrated set of parameters $(\bar{\alpha}, \bar{\beta}, \bar{\delta})$ for each asset and maturity.


\subsubsection{Calibration outcomes}

\begin{figure}[H]

    \centering
    \includegraphics[width=0.9\textwidth]{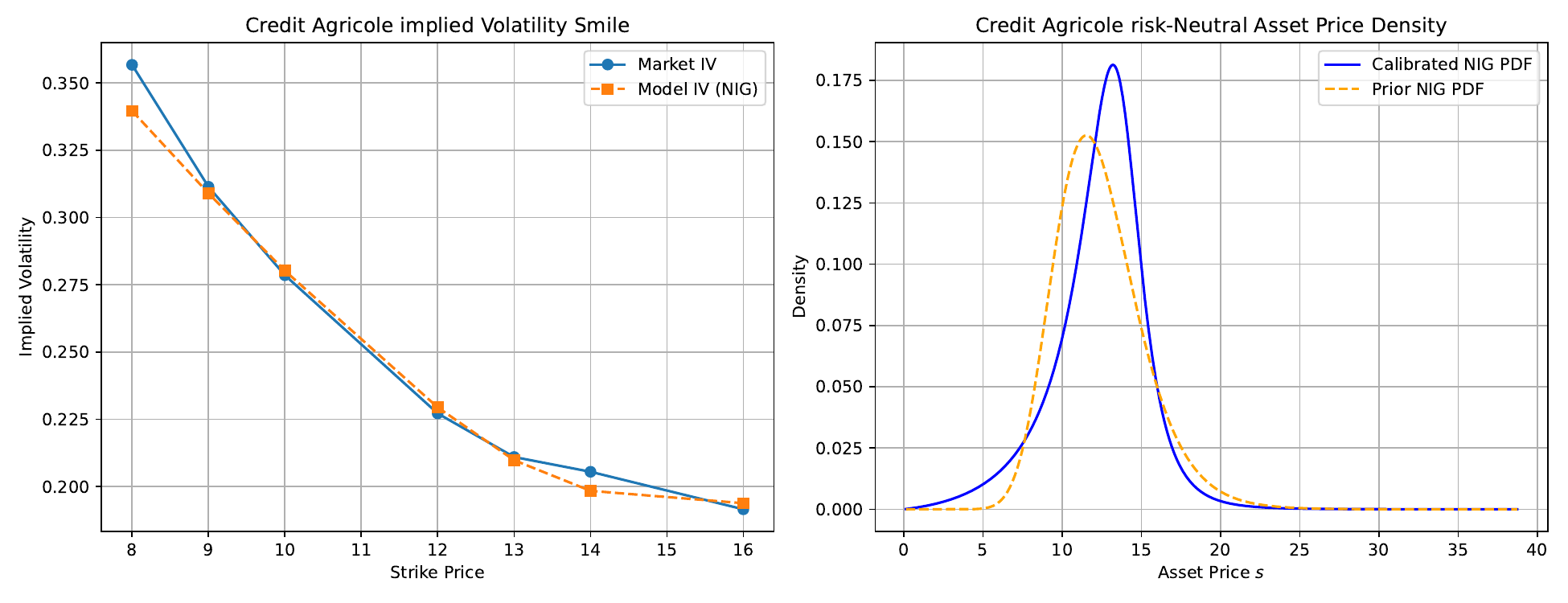} 
    \caption{\small Credit Agricole, 1-year expiry (19/12/2025), data as of 24/12/2024 with closing spot price at 12.91 EUR. Calibrated parameters: $\bar{\alpha} = 4.69, \bar{\beta} = -3.06, \bar{\delta} = 0.18$ with $\lambda = 5 \times 10^{-7}$. Left: market vs calibrated implied volatilities. Right: prior log-normal density function using ATM implied volatility vs calibrated density function.}
    \label{fig:CA_Calibrated_IV_PDF}
\end{figure}

\begin{figure}[H]
    \centering
    \includegraphics[width=0.9\textwidth]{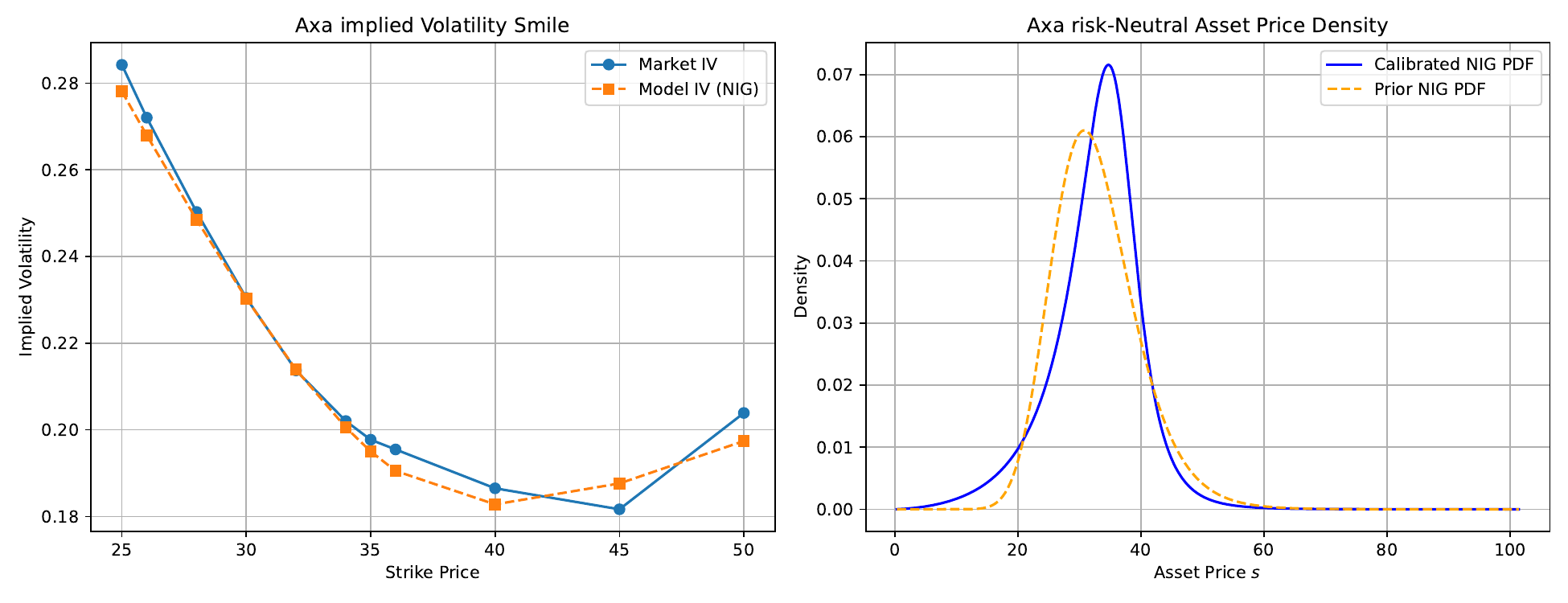} 
    \caption{\small AXA, 1-year expiry (19/12/2025), data as of 24/12/2024 with closing spot price at 33.8 EUR. Calibrated parameters: $\bar{\alpha} = 5.24, \bar{\beta} = -3.26, \bar{\delta} = 0.18$ with $\lambda = 5 \times 10^{-7}$. Left: market vs calibrated implied volatilities. Right: prior log-normal density function using ATM implied volatility vs calibrated density function.}
    \label{fig:Axa_Calibrated_IV_PDF}
\end{figure}

\begin{figure}[H]
    \centering
    \includegraphics[width=0.9\textwidth]{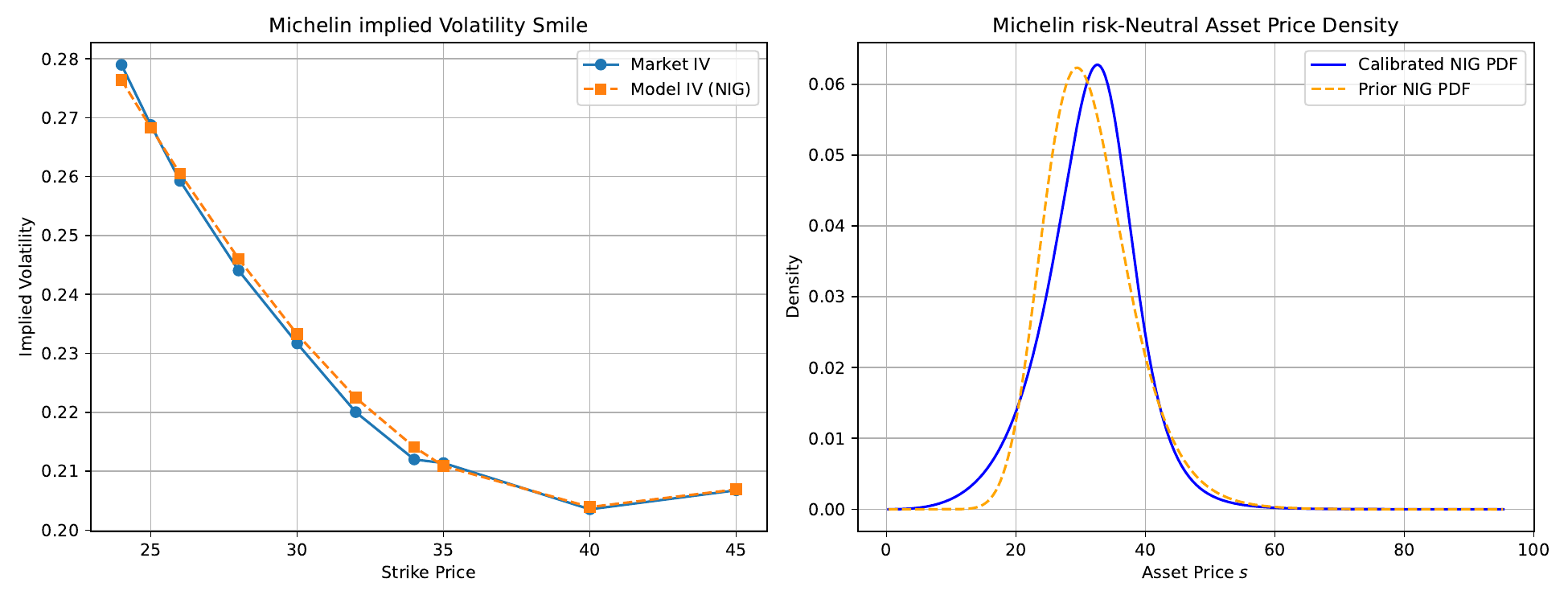} 
    \caption{\small Michelin, 1-year expiry (19/12/2025), data as of 24/12/2024 with closing spot price at 31.76 EUR. Calibrated parameters: $\bar{\alpha} = 6.2, \bar{\beta} = -3.31, \bar{\delta} = 0.26$ with $\lambda = 5 \times 10^{-7}$. Left: market vs calibrated implied volatilities. Right: prior log-normal density function using ATM implied volatility vs calibrated density function. }
    \label{fig:Michelin_Calibrated_IV_PDF}
\end{figure}

The calibrated NIG densities exhibit strong agreement with observed market option prices, capturing key features such as skew and smile. The resulting risk-neutral densities are arbitrage-free at each time slice. For Credit Agricole, the largest pricing error (normalized by the spot) is approximately 21 basis points, occurring in the tails. Around the ATM strike, errors are typically around 10 basis points (see Figure \ref{fig:CA_Calibrated_IV_PDF}). For AXA, the maximum discrepancy is about 10 basis points (see Figure \ref{fig:Axa_Calibrated_IV_PDF}), while for Michelin, it is approximately 6 basis points (see Figure \ref{fig:Michelin_Calibrated_IV_PDF}), indicating high calibration accuracy across all three assets. Each figure compares the calibrated NIG density with the prior lognormal distribution based on ATM implied volatility. As expected for equity markets, all distributions display fat tails and left skew, which the NIG model captures well. The inclusion of a Tikhonov regularization term stabilizes parameter estimates and prevents overfitting in regions with sparse or noisy quotes. Overall, the results confirm the NIG model’s ability to reflect empirical skewness and kurtosis, supporting its use in downstream tasks such as quantum-based pricing and copula-like multivariate modelling.

\section{Multidimensional option pricing using quantum computing}\label{sec:option_pricing_quantum}

We explore the power of quantum computing when addressing the problem of multidimensional option pricing.

\subsection{Multidimensional option pricing using copulas}\label{sec:multidim_option_pricing_copulas}
We are interested in multivariate option pricing of European-like options, where the payoff function can be written in general form as

\begin{equation*}
h\left(\mathbf{S}, K\right), \quad \mathbf{S} = \left(S_i(T),\, i = 1, 2, \ldots, N \right),
\end{equation*}
where, as usual, $h(\cdot)$ is a univariate payoff function that identifies the derivative contract, $S_i$ denotes the price of the $i^\text{th}$ underlying security, $T$ is the contract maturity and $K$ represents the contract strike. Below lists some common examples:
\begin{enumerate}
    \item Arithmetic basket call option,
    \begin{equation*}
    h\left(S_i(T), K\right) = \max\left( \frac{1}{N}\sum_{i=1}^{N} S_i(T) - K, \, 0 \right),
    \end{equation*}

    \item Worst-of put option,
    \begin{equation*}
    h\left(S_i(T), K\right) = \max\left( K - \min_{i = 1, \ldots, N} S_i(T), \, 0 \right),
    \end{equation*}
    
    \item Spread call option,
    \begin{equation*}
     h\left(S_1(T), S_2(T), K\right) = \max\left( S_1(T) - S_2(T) - K, \, 0 \right).
    \end{equation*}
\end{enumerate}

In general, given a multivariate payoff, the option value can be then formulated in terms of an expectation as
\begin{equation}\label{eq:multi_option_price}
    V(T, K) = e^{-rT} \mathbb{E}[h(\mathbf{S}, K)] = e^{-rT} \int_\Omega f(\mathbf{S}) h(\mathbf{S}, K) \, \D^N\mathbf{S},
\end{equation}
here written as well in integral form for convenience. In order to address the resolution of that integral via numerical techniques, the availability of the joint density function of the underlying assets is desired. In the following, a copula-based approach for deriving such joint density is described.

\subsubsection{Joint distribution via copulas}\label{sec:copula}

In the context of multivariate option pricing, especially when dealing with multiple underlying assets, it is essential to model the joint distribution of asset prices at maturity. While the marginal distributions of each asset can be independently inferred from market option prices (see Section \ref{sec:NIG_calibration}), their joint behaviour must account for inter-asset dependencies. 

\textit{Copulas} provide a powerful and flexible tool to model this dependence structure separately from the marginals. A copula is a multivariate distribution function defined on the unit cube $[0,1]^N$ with uniform marginals, which allows the construction of joint distributions from given marginals. More precisely, by using a copula, we can handle the individual univariate marginal distributions 
and their dependency separately, thanks to Sklar’s theorem, which guarantees the consistency between the copula-based 
joint distribution and each marginal distribution.

\begin{theorem}[Sklar’s Theorem \cite{nelsen2006copulas}]
For any joint distribution function $F$ on $\mathbb{R}^N$ of a random vector $\mathbf{X} = (X_1, \dots, X_N)$ with marginal distribution functions $F_1, \dots, F_N$, there exists a copula $C$ such that
\begin{equation} \label{eq:SklarThrFormula}
F(x_1, \dots, x_N) = \mathcal{C}(F_1(x_1), \dots, F_d(x_N)),
\end{equation}
for any $(x_1, \dots, x_N) \in \mathbb{R}^N$. If $F_1, \dots, F_N$ are continuous, then the copula $\mathcal{C}$ is unique. Conversely, for any marginal distributions $F_1, \dots, F_N$ and a $N$-variate copula $\mathcal{C}$, the function $F$ defined as in \eqref{eq:SklarThrFormula} is a valid joint distribution function with marginals $F_1, \dots, F_N$.
\end{theorem}

From formula \eqref{eq:SklarThrFormula}, if the marginals $F_1, \dots, F_N$ are differentiable with densities $f_1, \dots, f_N$, the copula $\mathcal{C}$ is differentiable with density $c$ given by

\begin{equation*}
c(u_1, \dots, u_N) = \frac{\partial^N}{\partial u_1 \cdots \partial u_N} \mathcal{C}(u_1, \dots, u_N),
\end{equation*}
then, with a direct derivation, the joint density $f$ of the vector $(X_1, \dots, X_N)$ can be written as:

\begin{equation}\label{eq:copula-density}
f(x_1, \dots, x_N) = c\left(F_1(x_1), \dots, F_N(x_N)\right) \cdot \prod_{i=1}^N f_i(x_i).
\end{equation}

This formula expresses the joint density as the product of two components:
\begin{itemize}
    \item The copula density evaluated at the marginal distribution functions, which captures the interdependence between variables;
    \item The product of the marginal densities, which captures the individual behaviour of each variable.
\end{itemize}

A popular choice of copula (and the one that will be considered in this work) is the Gaussian copula, due to its tractability. Let $\Phi_N(\cdot; \Sigma)$ be the $N$-dimensional standard normal cumulative distribution function with correlation matrix $\Sigma \in \mathbb{R}^{N \times N}$, and let $\Phi^{-1}$ denote the univariate standard normal quantile function. The Gaussian copula is defined as
\begin{equation*}
\mathcal{C}_\Sigma(u_1, \dots, u_N) = \Phi_N\left( \Phi^{-1}(u_1), \dots, \Phi^{-1}(u_N); \Sigma \right).
\end{equation*}
The corresponding copula density is
\begin{equation*}
c_\Sigma(u_1, \dots, u_N) = \frac{1}{\sqrt{\det \Sigma}} \exp\left( -\frac{1}{2} \mathbf{z}^\top (\Sigma^{-1} - \mathcal{I}) \mathbf{z} \right),
\end{equation*}
where $\mathbf{z} = (\Phi^{-1}(u_1), \dots, \Phi^{-1}(u_N))$ and $\mathcal{I}$ is the identity matrix.

\subsubsection*{Multidimensional option valuation with copulas}

Employing \eqref{eq:copula-density}, we can rewrite the pricing formulation in (\ref{eq:multi_option_price}) as 
\begin{equation}\label{eq:PricingwithCopula}
\begin{aligned}
V(T, K) &= e^{-rT} \E[h(\mathbf{S}, K)] \\
        &= e^{-rT} \int _\Omega  h(\mathbf{S}, K) c\left(F_1(S_1), \dots, F_N(S_N)\right) \cdot \prod_{i=1}^N f_i(S_i) \D^N \mathbf{S} \\
        &= e^{-rT} \E^{\mathrm{ind}}[h(\mathbf{S}, K) c\left(F_1(S_1), \dots, F_N(S_N)\right)],
\end{aligned}
\end{equation}
where, again, $\mathbf{S}$ is the vector of random variables representing the asset prices at expiry $T$, $h(x)$ is the final payoff function and $\E^{ind}[\cdot]$ is the expectation operator applied to $\mathbf{S}$ by considering its components $S_i$ as independent. By using \eqref{eq:PricingwithCopula}, the pricing of a payoff under a \emph{correlated} joint distribution can be rewritten as an expectation under \emph{independent} marginals, at the cost of weighting the payoff by the copula density \(c(F_1(\cdot),\dots,F_N(\cdot))\). In other words, correlation is entirely captured by this multiplicative weight.

Note then that, in order to build (and work with) the copula approach described above, both the marginal density and distribution functions are required. In the derivatives pricing framework, it is often the case that no analytical closed-form for such expressions are available, or their tractability is not efficient in numerical and/or computational terms. As example, for the NIG model presented in Section \ref{sec:exponential_NIG_model}, although the density function is known, given by \eqref{eq:NIG_density}, the distribution function (and its inverse, the quantile function) needs to be treated numerically, typically incurring in high computational costs and instabilities. Thus, to make our approach generally applicable (and open the door for the utilisation of quantum algorithms which can potentially provide remarkable computational benefits), in next Section \ref{sec:cos_density_estimation}, a non-parametric density estimation relying on cosine basis functions is proposed, recalling its main theoretical properties, which will be later used to theoretically prove the quantum advantage.

\subsubsection{Cosine-series density (and distribution) estimation}\label{sec:cos_density_estimation}

Let \(f:\mathbb{R}\to\mathbb{R}_{\ge 0}\) be a probability density function supported (or effectively supported) on a finite interval \([a,b]\), associated to a random variable $X$.
In orthogonal-series density estimation, the target density is expanded in a complete orthonormal basis of functions on \([a,b]\),
and its coefficients are obtained by projection under the \(L^{2}\) inner product.
For the cosine basis
\begin{equation*}
\gamma_k(x)=
\begin{cases}
\displaystyle \frac{1}{\sqrt{b-a}}, & k=0, \\[8pt]
\displaystyle \sqrt{\frac{2}{b-a}}\,
\cos\!\Big(\frac{k\pi(x-a)}{b-a}\Big), & k\ge1,
\end{cases}
\end{equation*}
we have the orthonormality property
\begin{equation*}
\int_a^b \gamma_k(x)\,\gamma_\ell(x)\,\D x = \delta_{k\ell},
\qquad k,\ell\ge0,
\end{equation*}
with each basis function uniformly bounded by
\begin{equation}\label{eq:GammaFuncBound}
|\gamma_k(x)| \le \sqrt{\frac{2}{b-a}}, \qquad x\in[a,b],\ k\ge0,
\end{equation}
so that any square-integrable function \(f\in L^2([a,b])\) admits the cosine expansion
\begin{equation}\label{eq:cosine_expansion}
f(x) = \sum_{k=0}^{\infty} a_k\, \gamma_k(x),
\end{equation}
where
\begin{equation}\label{eq:cosine_coef}
a_k := \mathbb{E}[\gamma_k(X)] = \int_a^b f(x)\, \gamma_k(x)\,\D x.
\end{equation}

\begin{remark}
    Sometimes, to obtain the $a_k$ coefficients, it might be convenient to work with strictly positive basis functions. In that case, the following transformation can be applied,
    \begin{equation*}
        \gamma^+_k(x) = \frac{1}{2} + \frac{1}{2 } \sqrt{\frac{b-a}{2}} \gamma_k(x),    
    \end{equation*}
    which satisfies \( 0 \le \gamma^+_k(x) \le 1 \). Then, the cosine coefficients can be equivalently obtained by
    \begin{equation}\label{eq:ak_def}
        a_k := \mathbb{E}[\gamma_k(X)] = \sqrt{\frac{2}{b-a}}\left( 2 \mathbb{E}[\gamma^+_k(X)] - 1 \right).
    \end{equation}
    
\end{remark}

Truncating the series in \eqref{eq:cosine_expansion} to \(\mathcal{K}\) terms yields the approximation
\begin{equation*}
f(x) \approx f^\mathcal{K}(x) := \sum_{k=0}^{\mathcal{K}-1} a_k\,\gamma_k(x),
\end{equation*}
which forms the basis of the Fourier–cosine (COS) method, widely used for density and option pricing computations (see e.g \cite{FangOosterlee2008}).
The convergence of \(f^\mathcal{K}\) to \(f\) depends on the smoothness or analyticity of \(f\),
as established in the following theorem.

\begin{theorem}[Uniform cosine--series approximation on a finite interval]
\label{th:CosSeriesProxyThm}
Let \(f:[a,b]\to\mathbb{R}\) be a real-valued function, and define its cosine coefficients
\begin{equation*}
a_0 := \frac{1}{b-a}\int_a^b f(x)\,\D x, \qquad
a_k := \frac{2}{b-a}\int_a^b f(x)\cos\!\left(\frac{k\pi(x-a)}{b-a}\right)\,\D x,\quad k\ge1.
\end{equation*}
Define the \(\mathcal{K}\)-term partial sum
\begin{equation*}
f^\mathcal{K}(x) := \sum_{k=0}^{\mathcal{K}-1} a_k \cos\!\left(\frac{k\pi(x-a)}{b-a}\right), 
\qquad x\in[a,b].
\end{equation*}

Assume that the truncation interval \([a,b]\) is chosen such that \(f\) and its first \(m-1\) derivatives vanish (or are negligible) at the endpoints:
\begin{equation*}
f^{(j)}(a)=f^{(j)}(b)=0, \qquad j=0,\ldots,m-1.
\end{equation*}
This condition is satisfied, in practice, when \(f\) is smooth and rapidly decaying outside \([a,b]\).

Then:
\begin{enumerate} 
\item \textbf{(Algebraic case)}  
If \(f\in C^{m}([a,b])\) and \(f^{(m)}\) has bounded variation on \([a,b]\), there exists \(\zeta^{\mathrm{alg}}>0\) such that, for every \(\mathcal{K} \ge 1\), 
\begin{equation*}
\sup_{x\in[a,b]} |f(x)-f^\mathcal{K}(x)| \le \zeta^{\mathrm{alg}} \mathcal{K}^{-m},
\end{equation*}
and equivalently, the coefficients satisfy \( |a_k| = O(k^{-(m+1)}) \).

\item \textbf{(Exponential case)}  
If \(f\) extends analytically to the complex strip
\begin{equation*}
\{\,z\in\mathbb{C} : |\Im z|<\rho\,\}
\end{equation*}
containing \([a,b]\) for some \(\rho>0\), then there exist constants \(\zeta^{\mathrm{exp}}, \nu>0\) such that, for every \(\mathcal{K} \ge 1\),
\begin{equation*}
\sup_{x\in[a,b]} |f(x)-f^\mathcal{K}(x)| \le \zeta^{\mathrm{exp}} e^{-\nu \mathcal{K}},
\end{equation*}
i.e.\ the cosine expansion converges uniformly at an exponential rate.
\end{enumerate}

The constants \(\zeta^{\mathrm{alg}}, \zeta^{\mathrm{exp}}, \nu\) depend on \(f\), the interval \([a,b]\), and the regularity parameters, but not on \(\mathcal{K}\).
\end{theorem}

\begin{remark}
The endpoint assumption effectively enforces a compactly supported of \(f\) on \([a,b]\), ensuring the boundary terms vanish in repeated integration by parts.  This is standard in Fourier--cosine and spectral approximation theory; see, e.g., 
Boyd~\cite{Boyd1989}, 
Trefethen~\cite{Trefethen2000}, 
Zygmund~\cite{Zygmund1959}, 
and Fang and Oosterlee~\cite{FangOosterlee2008}.
\end{remark}

\subsubsection{Estimating the marginal distributions}\label{sec:marginals}

Next, we explain the process to estimate the marginal cumulative distribution functions. Here, we only consider sufficiently smooth marginal distributions that can be well approximated by cosine series (see the assumptions in Theorem \ref{th:CosSeriesProxyThm}), which is usually the case in option pricing.

Let $X_i$ denote the marginal random variable, whose corresponding density and distribution functions are $f_i$ and $F_i$, respectively. Then, given estimated coefficients \( \hat{a}^{X_i}_k \approx a^{X_i}_k := \mathbb{E}[\gamma_k(X_i)]\) (see Equation \eqref{eq:cosine_coef}), we have a cosine series approximating \( f_i \) as
\begin{equation}\label{eq:fi_hat}
    f_i \approx \hat{f}_i(x) := \sum_{k=0}^{\mathcal{K}_i-1} \hat{a}^{X_i}_k \, \gamma_k(x).
\end{equation}
Furthermore, by integrating this, we get an approximation \( \hat{F}_i \) for \( F_i \). In fact, since the accuracy of the cosine series approximation is guaranteed not on the entire real axis but in a finite interval, we set \( \hat{F}_i \) to 0 or 1 outside the interval. Namely, we define
\begin{equation} \label{eq:Fi_hat}
    \hat{F}_i(x) :=
    \begin{cases}
        0, & x < a_i, \\
        \displaystyle \sum_{k=0}^{\mathcal{K}_i} \hat{a}^{X_i}_k \, \Gamma_{k,[a_i, b_i]}(x), & a_i \le x < b_i, \\
        1, & x \ge b_i,
    \end{cases}
\end{equation}
where,
\begin{equation*}
    \Gamma_{k,[a_i, b_i]}(x) := \int_{a_i}^{x} \gamma_k(t) \, \D t,
\end{equation*}
is given by
\begin{equation*}
\Gamma_{k,[a_i,b_i]}(x) =
\begin{cases}
\dfrac{x-a_i}{\sqrt{\,b_i-a_i\,}}, & k=0,\\[10pt]
\dfrac{\sqrt{2(b_i-a_i)}}{k\pi}\,
\sin\!\displaystyle\Big(\dfrac{k\pi(x-a_i)}{b_i-a_i}\Big), & k\ge 1.
\end{cases}
\end{equation*}




\subsection{Quantum algorithm for multidimensional options pricing}\label{sec:quantum_multidim}

In this section, a quantum computing-based approach to address the problem of multidimensional option valuation formulated above is proposed, discussing both theoretical and practical implications. We begin summarizing the employed quantum routine, followed by some theoretical results supporting the quantum advantage, which will be empirically confirmed by the experiments in the next Section \ref{sec:quantum_experimental}.

\subsubsection{Quantum Accelerated Monte Carlo techniques}\label{sec:QAMC}

The Monte Carlo methods are well-known integration techniques for solving option pricing problems, when formulated in terms of expectations. This method gives an approximation of the value of definite integrals by generating random samples within the integration region and computing the average value of the function evaluated in these samples \cite{Glasserman2004}.

Let us consider the computation of an expectation of a function of interest $\phi$ (for example, the payoff $h$ in the options pricing problem described in Section \ref{sec:multidim_option_pricing_copulas} or the cosine basis functions $\gamma_k$ as in Section \ref{sec:cos_density_estimation}) acting on a multidimensional random variable $\mathbf{X}$, given in the form of a $N$-dimensional definite integral, namely,
\begin{equation*}
\mathbb{E}[\phi(\mathbf{X})] = \int_\Omega f(\mathbf{x}) \phi(\mathbf{x})\,\D^N \mathbf{x},
\end{equation*}
where $f$ is a density with compact support $\Omega$. Note that the definitions of both the price of a multidimensional option in \eqref{eq:multi_option_price} and the cosine series expansion coefficients in \eqref{eq:cosine_coef} can be cast into this formulation.

Thus, the well-established CMC method consists in generating $L$ independent and identically distributed $N$-dimensional samples $\mathbf{X}_l$, for $l = 0, \dots, L-1$, drawn from the distribution associated with $f$, such that the value of the integral is approximated by 
\begin{equation}\label{eq:MonteCarlo}
\int_\Omega f(\mathbf{x})\phi(\mathbf{x})\,\D^N\mathbf{x} \approx \frac{1}{L}\sum_{l=0}^{L - 1} \phi(\mathbf{X}_l).
\end{equation}

Since this method can be computationally demanding for certain types of integrals, in recent years the advantages offered by quantum computing have been exploited to develop QAMC techniques \cite{Montanaro2015, Rebentrost_2018, Gomez_2022}, which promise a quadratic improvement, in terms of the estimation error, in the number of queries required compared to its classical counterpart. 

The common starting point relies on a discrete version of the integral, namely a Riemann sum, defined in $J = 2^{Nn}$ discrete points, being $n$ the number of qubits employed in the discretization for each dimension\footnote{We have assumed, without any loss of generality, the same number of discrete points in every space direction.}, which is given by
\begin{equation}\label{eq:riemann}
\int_\Omega f(\mathbf{x})\phi(\mathbf{x})\,\D^N\mathbf{x} \approx \sum_{j=0}^{J - 1} f(\mathbf{x}_j)\phi(\mathbf{x}_j).
\end{equation}

The idea behind the QAMC method is to encapsulate the value of the integral within the amplitudes of a quantum state, and then maximize the probability of obtaining this value when performing a measurement. For this purpose, we then assume that the following state on a circuit of $Nn+1$ qubits can be constructed\footnote{We have intentionally omitted the normalization constants for the sake of clarity.}
\begin{equation}\label{eq:U}
\begin{aligned}
	|\psi\rangle = \mathcal{U}|x_1\rangle^n \dots |x_N\rangle^n |0\rangle &= \sum_{j=0}^{2^{Nn}-1} \sqrt{f(\mathbf{x}_j)\phi(\mathbf{x}_j)} \, |x_1\rangle^n \dots |x_N\rangle^n |1\rangle \\
    &+ \sum_{j=0}^{2^{Nn}-1} \sqrt{1 - f(\mathbf{x}_j)\phi(\mathbf{x}_j)} \, |x_1\rangle^n \dots |x_N\rangle^n |0\rangle,
\end{aligned}
\end{equation}
where $\mathcal{U}$ is a quantum operator which encapsulates the (square root of the) Riemann sum that approximates the desired integral into the amplitude of the ancillary qubit's state $|1\rangle$. The oracle $\mathcal{U}$ is typically composed of two operators, one loading the density, $f$, and one loading the function of interest, $\phi$. There exist many methods in the literature to perform this type of quantum state preparation, see e.g. \cite{grover2002creatingsuperpositionscorrespondefficiently, Zoufal2019, Stamatopoulos2020, Vazquez2020, Moosa2023}, for which, as considered here, an auxiliary ancilla qubit is typically required (the last qubit of the quantum state $|\psi\rangle$ in \eqref{eq:U}).

Then, the value of the integral can be estimated through the QAE routine \cite{Brassard2002, Montanaro2015}, a quantum algorithm that allows to efficiently retrieve the amplitude information from a quantum state. In this particular formulation (where a square root encoding is employed), the probability of obtaining $|1\rangle$ when measuring the state $|\psi\rangle$, i.e., the squared amplitude of the state, is precisely the Riemann estimator of the integral in \eqref{eq:riemann} that, in turn, approximates $\mathbb{E}[\phi(\mathbf{x})]$. As it will be shown in the following Sections \ref{sec:quantum_theory} and \ref{sec:quantum_experimental}, the application of the QAE to the proposed multivariate option valuation methodology results in remarkable accelerations with respect to the CMC approaches, from both theoretical and empirical viewpoints.

\subsubsection{Quantum advantage: theoretical results}\label{sec:quantum_theory}

Let us first recall the well-known result that forms the basis of the QAMC method, namely the QAE routine, formulated in the next theorem.

\begin{theorem}[Quantum Amplitude Estimation; Theorem 2.3 in \cite{Montanaro2015}]
\label{th:qmci}
Let $\varrho, \epsilon \in (0,1)$. 
Assuming we have access to the state preparation oracle $\mathcal{A}_Y$ for a random variable $Y \in \mathbb{R}^N$ and the controlled rotation oracle $W_\phi$ for a function $\phi : \mathbb{R}^N \to [0,1]$, there exists a quantum algorithm that, with probability at least $1 - \varrho$, outputs an $\epsilon$-approximation of $\mathbb{E}_Y[\phi(Y)]$, querying $\mathcal{A}_Y$ and $W_\phi$ 
\begin{equation*}
O\!\left(
\frac{1}{\epsilon} 
\log\!\frac{1}{\varrho}
\right)
\end{equation*}
times each.
\end{theorem}

\begin{remark}
    Note that, in this context, we assume that querying a quantum oracle one time is equivalent to draw a single sample from a given distribution in the classical computation, avoiding as well the discussion on any particular computational capability aspect or technology readiness. Then, in Section \ref{sec:quantum_experimental}, the number of samples and queries will be fairly compared under these premises.
\end{remark}

\subsubsection*{Quantum estimation of the marginal distributions}

Before deriving a rigorous theoretical result for the convergence in estimating the marginal distributions with QAMC techniques, we briefly discuss how to build the quantum state required to apply them (see Section \ref{sec:QAMC}). First, to follow the approach described in Sections \ref{sec:cos_density_estimation} and \ref{sec:marginals}, we need to define an oracle that encapsulates the Riemann sum approximating the expectation in \eqref{eq:cosine_coef}. For that, as it is common in the literature, two quantum operations are combined, loading the probability density function and a function of interest into the amplitude of a quantum state. In this case, to load $f_i$, we assume the availability of an oracle $\mathcal{A}_{X_i}$ such that\footnote{As we are treating with marginal distributions, the dimensionality of the problem is one, allowing us to avoid the vector-like bold notation.}
\begin{equation}\label{eq:AXi}
    \mathcal{A}_{X_i}|0\rangle^n = \sum_{j=0}^{2^{n}-1} \sqrt{f_i(x_j)} \, |x\rangle^n.
\end{equation}
Next, by employing a controlled rotation, the function of interest is loaded. Then, let us assume that we have access to the controlled rotation operation $W_{\gamma_k}$, for $k = 0, \ldots, \mathcal{K}_i - 1$, which transforms state \eqref{eq:AXi} into the state
\begin{equation}\label{eq:Uak}
\begin{aligned}
    \mathcal{U}_{a_k}|0\rangle^n|0\rangle := W_{\gamma_k}(A_{X_i} \otimes \mathcal{I})|0\rangle^n|0\rangle &= \sum_{j=0}^{2^{n}-1} \sqrt{f_i(x_j)\gamma_k(x_j)} \, |x\rangle^n |1\rangle \\
    &+ \sum_{j=0}^{2^{n}-1} \sqrt{(1 - f_i(x_j)\gamma_k(x_j)} \, |x\rangle^n |0\rangle,
\end{aligned}
\end{equation}
where $\mathcal{I}$ is the identity matrix and we have defined the operator $\mathcal{U}_{a_k}:= W_{\gamma_k}(A_{X_i} \otimes \mathcal{I})$, which fit into the general description of QAMC in Section \ref{sec:QAMC}. Below, the main theorem providing the accuracy of the QAMC method and the costs associated when applied it to recover the marginal distributions is presented.

 \begin{theorem}[Quantum complexity recovering the marginal distributions]
 \label{th:estimation_Fi}
Let $\varrho_i, \epsilon_i \in (0,1)$. Let $X_i$ be a real-valued random variable following the distribution $f_i$. 
Assume the following:
\begin{enumerate}
    \item $f_i$ has the properties of Theorem \ref{th:CosSeriesProxyThm}.
    \item Access to the oracle $\mathcal{U}_{a_k}$ as in \eqref{eq:Uak} for $k = 0, \ldots, \mathcal{K}_i - 1$. 
    \item For some $[a_i, b_i] $, $F_i(a_i) \le \epsilon_i/2$ and $F_i(b_i) \ge 1 - \epsilon_i$ hold. \label{it:a_ib_iF_i}
\end{enumerate}
Then, with probability at least $1 - \varrho_i$, we get $\hat{F}_i$ such that
\begin{equation}\label{eq:FiHat_Fi_Errors}
\left| \hat{F}_i(x) - F_i(x) \right| \le \epsilon_i
\end{equation}
for any $x \in \mathbb{R}$ by Theorem \ref{th:qmci}, querying $\mathcal{U}_{a_k}$ 
for $k = 0, \ldots, \mathcal{K}_i - 1$
\begin{equation*}
O\!\left(
\frac{ \sqrt{b_i-a_i} \mathcal{K}^2_i}{\epsilon_i}
\log\!\frac{\mathcal{K}_i}{\varrho_i}
\right)
\end{equation*}
times. 

\begin{itemize}

\item For the algebraic case,  we set
\begin{equation}\label{eq:alg_KiFormula}
\mathcal{K}_i =
\left\lceil
\left(
\frac{4 \zeta^{\mathrm{alg}}_{i} (b_i - a_i)}{\epsilon_i}
\right)^{\frac{1}{m_i}}
\right\rceil,
\end{equation}
where $\zeta^{\mathrm{alg}}_{i}$ is a real number such that
\begin{equation}\label{eq:alg_uniformBoundfi}
\sup_{x \in [a_i, b_i]}
\left| f_i^{\mathcal{K}}(x) - f_i(x) \right|
\le
\zeta^{\mathrm{alg}}_{i} \mathcal{K}^{-m_i}
\end{equation}
holds for any $\mathcal{K} \in \mathbb{N}$. 

\item For the exponential case,  we set
\begin{equation}\label{eq:exp_KiFormula}
\mathcal{K}_i =
\left\lceil
\log \left[ \left(
\frac{4 \zeta^{\mathrm{exp}}_{i} (b_i - a_i)}{\epsilon_i}
\right)^{\frac{1}{\nu_i}} \right] 
\right\rceil,
\end{equation}
where $\zeta^{\mathrm{exp}}_{i}$ is a real number such that
\begin{equation}\label{eq:exp_uniformBoundfi}
\sup_{x \in [a_i, b_i]}
\left| f_i^{\mathcal{K}}(x) - f_i(x) \right|
\le
\zeta^{\mathrm{exp}}_{i} e^{-\nu_i \mathcal{K}}
\end{equation}
holds for any $\mathcal{K} \in \mathbb{N}$.

\end{itemize}

\end{theorem}

\begin{proof}
Because of the definition of $\hat{F}_i$ in \eqref{eq:Fi_hat} and Assumption \ref{it:a_ib_iF_i}, it is immediately seen that \eqref{eq:FiHat_Fi_Errors} holds for any $x \in (-\infty, -a_i] \cup [b_i, \infty)$. 
Thus, we hereafter focus on the case that $x \in (a_i,b_i)$.
We start by evaluating $|\hat{f}_i(x) - f_i(x)|$. Decomposing it as
\begin{equation} \label{eq:hatfi-fiDecomposition}
|\hat{f}_i(x) - f_i(x)| \le |\hat{f}_i(x) - f_i^{\mathcal{K}_i}(x)| + |f_i^{\mathcal{K}_i}(x) - f_i(x)|
\end{equation}

The first term is the Monte Carlo error and the second one the series truncation error.
We bound each term separately. For the second term, for $x \in (a_i, b_i)$, in the algebraic convergence case, we have
\begin{equation}\label{eq:fifKiBoundAlg}
 \left| f_i^{\mathcal{K}_i}(x) - f_i(x) \right|
\le
\zeta^{\mathrm{alg}}_{i} \mathcal{K}_i^{-m_i}
\le
\frac{\epsilon_i}{4 (b_i - a_i)}, 
\end{equation}
where we use \eqref{eq:alg_uniformBoundfi} with $\mathcal{K}_i$ defined as in \eqref{eq:alg_KiFormula}. In the exponential convergence case, we get
\begin{equation}\label{eq:fifKiBoundExp}
 \left| f_i^{\mathcal{K}_i}(x) - f_i(x) \right|
\le
\zeta^{\mathrm{exp}}_{i} e^{-\nu_i \mathcal{K}_i}
\le
\frac{\epsilon_i}{4 (b_i - a_i)}, 
\end{equation}
where we use \eqref{eq:exp_uniformBoundfi} with $\mathcal{K}_i$ defined as in \eqref{eq:exp_KiFormula}. To bound the first one, we temporarily assume that,
relying on Theorem \ref{th:qmci} with $\delta = \frac{\varrho_i}{\mathcal{K}_{i}}$ and $\epsilon = \frac{\epsilon_i}{4 \mathcal{K}_{i} \sqrt{2(b_i - a_i)}}$, a quantum algorithm outputs the estimation $\widehat{\mathbb{E}}_{X_i}[\gamma_k(X_i)]$ for every $k = 0,...,\mathcal{K}_{i}-1$ such that
\begin{equation} \label{eq:errorExpGammaHatXi}
\left|
\widehat{\mathbb{E}}_{X_i}[\gamma_k(X_i)] - \mathbb{E}_{X_i}[\gamma_k(X_i)]
\right| = |\hat{a}^{X_i}_{k} - a^{X_i}_{k}|
\le \frac{\epsilon_i}{4 \mathcal{K}_{i} \sqrt{2(b_i - a_i)}}.
\end{equation}
We then have
\begin{equation}\label{eq:fiMinusfiKiBound}
\begin{aligned}
\left| \hat{f}_i(x) - f_i^{\mathcal{K}_i}(x) \right|
& = \left|
\sum_{k=0}^{\mathcal{K}_{i}-1}
 \left(\hat{a}_k^{X_i} - a_k^{X_i}\right) \gamma_k(x) \right|  \\
& \le
\sum_{k=0}^{\mathcal{K}_{i}-1}
\left| \hat{a}_k^{X_i} - a_k^{X_i} \right| \sqrt{\frac{2}{b_i-a_i}} \\
& \le \frac{\epsilon_i}{4 (b_i-a_i)} 
\end{aligned}
\end{equation}
where we use \eqref{eq:GammaFuncBound} at the first inequality and \eqref{eq:errorExpGammaHatXi} at the second inequality.

Combining \eqref{eq:hatfi-fiDecomposition}, \eqref{eq:fifKiBoundAlg} (or \eqref{eq:fifKiBoundExp}) and \eqref{eq:fiMinusfiKiBound} gives
\begin{equation*}
\left| \hat{f}_i(x) - f_i(x) \right| \le \frac{\epsilon_i}{2 (b_i - a_i)}.
\end{equation*}

Integrating this over $(a_i, x]$ yields
\begin{equation*}
\begin{aligned}
\left| F_i(x) - \hat{F}_i(x) \right|
&\le
|F_i(a_i)|
+
\int_{a_i}^x
\left| \hat{f}_i(y) - f_i(y) \right| \D y \\
&\le
\frac{\epsilon_i}{2} + \frac{\epsilon_i}{2 (b_i - a_i)} (x-a_i) \\
&\le \epsilon_i
\end{aligned}
\end{equation*}
for $x \in (a_i, b_i)$.

To complete the proof, let us prove the statements on the success probability and complexity.
Since the probability that each of 
the applications of Theorem \ref{th:qmci} with $\varrho = \frac{\varrho_i}{\mathcal{K}_{i}}$ and $\epsilon = \frac{\epsilon_i}{4 \mathcal{K}_{i} \sqrt{2(b_i - a_i)}}$ for every $k = 0,...,\mathcal{K}_{i}-1$ outputs $\widehat{\mathbb{E}}_{X_i}[\gamma_k(X_i)]$ satisfying (\ref{eq:errorExpGammaHatXi}) is at least $1 - \frac{\varrho_i}{\mathcal{K}_{i}}$,
the probability that \emph{all} of them output such estimations is
\begin{equation*}
\prod_{k=0}^{\mathcal{K}_{i} - 1} \left(1 - \frac{\varrho_i}{\mathcal{K}_{i}}\right)
= \left(1 - \frac{\varrho_i}{\mathcal{K}_{i}}\right)^{\mathcal{K}_{i}}.
\end{equation*}

Using the inequality $(1 - x)^p \ge 1 - px$ for $x \in [0,1]$ and integer $p \in \N$, we obtain
\begin{equation*}
\left(1 - \frac{\varrho_i}{\mathcal{K}_{i}}\right)^{\mathcal{K}_{i}}
\ge 1 - \varrho_i.    
\end{equation*}

From Theorem \ref{th:qmci}, the number of queries to each $\mathcal{U}_{a_k}$ (composed of the state preparation oracle $\mathcal{A}_{X_i}$ and the controlled rotation oracle $W_{\gamma_k}$, see \eqref{eq:Uak}) is
\begin{equation*}
O\!\left( \frac{ \sqrt{b_i-a_i} \mathcal{K}_{i}}{\epsilon_i} \log\!\left( \frac{\mathcal{K}_{i}}{\varrho_i} \right) \right).
\end{equation*}

Finally, summing them up for $k = 0,...,\mathcal{K}_{i} - 1$, we get 
\begin{equation*}
O\!\left( \frac{ \sqrt{b_i-a_i} \mathcal{K}_{i}^2}{\epsilon_i} \log\!\left( \frac{\mathcal{K}_{i}}{\varrho_i} \right) \right).
\end{equation*}

\end{proof}

\subsubsection*{Quantum estimation of the multidimensional option price}

Once the marginal distribution functions are computed (and having defined a copula), the multidimensional option pricing machinery described in Section \ref{sec:multidim_option_pricing_copulas} can be readily applied, where the QAMC methods can be further employed, specifically, in the computation of the integral/expectation in either expression \eqref{eq:multi_option_price} or expression \eqref{eq:PricingwithCopula}. Next, we describe how to proceed to encapsulate the required quantum state in each case which, from now on, we termed as \textit{joint} and \textit{independent} formulations, respectively.

For the joint case, we need to build a quantum state that resembles the integral value in \eqref{eq:multi_option_price} using an approximated joint density obtained via copulas as
\begin{equation*}
\hat{f}(\mathbf{x}) = \hat{f}(x_1, \dots, x_N) := c\left(\hat{F}_1(x_1), \dots, \hat{F}_N(x_N)\right) \cdot \prod_{i=1}^N \hat{f}_i(x_i),
\end{equation*}
where $\hat{f}_i$ and $\hat{F}_i$, $i=1,\dots,N$, are approximated density and distribution functions given by \eqref{eq:fi_hat} and \eqref{eq:Fi_hat}, respectively. Then, let us now assume the access to a quantum operator $\mathcal{A}_{\mathbf{X}}$ which acts on an initial zero state as
\begin{equation}\label{eq:AX}
    \mathcal{A}_{\mathbf{X}} |0\rangle^n \dots |0\rangle^n = \sum_{j=0}^{2^{Nn}-1} \sqrt{\hat{f}(\mathbf{x}_j)} \, |x_1\rangle^n \dots |x_N\rangle^n,
\end{equation}
so it loads the square root of the approximated joint density function $\hat{f}$ in the amplitude of a quantum state. Next, to approximate the option price, we then need to load the payoff function $h(\cdot)$, for which, an oracle $W_h$ is required, after whose application to the previously defined state \eqref{eq:AX}, we obtain
\begin{equation}\label{eq:UV}
\begin{aligned}
    W_h(\mathcal{A}_{\mathbf{X}} \otimes \mathcal{I})|0\rangle^n \dots |0\rangle^n |0\rangle &= \sum_{j=0}^{2^{Nn}-1} \sqrt{\hat{f}(\mathbf{x}_j)h(\mathbf{x}_j)} \, |x_1\rangle^n \dots |x_N\rangle^n |1\rangle \\
    &+ \sum_{j=0}^{2^{Nn}-1} \sqrt{(1 - \hat{f}(\mathbf{x}_j)h(\mathbf{x}_j)} \, |x_1\rangle^n \dots |x_N\rangle^n |0\rangle.
\end{aligned}
\end{equation}

In order to cast these derivations into the general description of QAMC from Section \ref{sec:QAMC}, we denote by $\mathcal{U}_V :=  W_h(\mathcal{A}_{\mathbf{X}} \otimes \mathcal{I})$ the oracle that constructs a quantum state which allows to estimate the multidimendional option price emplyoing the joint formulation. Under the premises of Theorem \ref{th:qmci}, a QAMC algorithm provides an approximation of the (non-discounted) price given by the $\E[h(\mathbf{X})]$ with precision less than a prescribed $\epsilon_V$, within a given confidence $1-\varrho_V$, and querying $\mathcal{U}_V$ an order $O\left(\frac{1}{\epsilon_V}\log\frac{1}{\varrho_V}\right)$ of times.

The independent case leverages the decomposition, thanks to the copula properties, of the joint density in two terms as shown in \eqref{eq:copula-density}, and already exploited in \eqref{eq:PricingwithCopula}.
Again, given the approximations of the distribution functions $\widehat{F}_1, \ldots, \widehat{F}_N$ in the form of \eqref{eq:Fi_hat}, we define an adjusted payoff function as
\begin{equation}\label{PayoffCopulaAdjusted}
\hat{H}(\mathbf{x}) = \hat{H}(x_1, \dots, x_N) := \frac{1}{c_{\max}}\, h(x)\, c\!\left( \hat F_1(x_1), \ldots, \hat F_N(x_N) \right),  \quad c_{\max} := \max_{u \in [0,1]^N} c(u), 
\end{equation}
which is an approximation of
\begin{equation}\label{TruePayoffCopulaAdjusted}
H(\mathbf{x}) = \frac{1}{c_{\max}}\, h(x)\, c\!\left( F_1(x_1), \ldots, F_N(x_N) \right).
\end{equation}

Next, an oracle, $\mathcal{A}^{\mathrm{ind}}_{\mathbf{X}}$, encapsulating the \emph{independent} joint distribution, $f^{\mathrm{ind}}(\mathbf{x}) = f^{\mathrm{ind}}(x_1, \dots, x_N) :=\prod_{i=1}^N f_i(x_i)$ is required. Thus, let us assume that we can build, as before, a $Nn$-qubit quantum state as
\begin{equation}\label{eq:AX_ind}
    \mathcal{A}^{\mathrm{ind}}_{\mathbf{X}} |0\rangle^n \dots |0\rangle^n = \sum_{j=0}^{2^{Nn}-1} \sqrt{f^{\mathrm{ind}}(\mathbf{x}_j)} \, |x_1\rangle^n \dots |x_N\rangle^n.
\end{equation}
In order to load the (approximated) adjusted payoff function from \eqref{PayoffCopulaAdjusted}, we again consider a quantum rotation operator $W_{\hat{H}}$ such that
\begin{equation*}
\begin{aligned}
    W_{\hat{H}}(\mathcal{A}^{\mathrm{ind}}_{\mathbf{X}} \otimes \mathcal{I})|0\rangle^n \dots |0\rangle^n |0\rangle &= \sum_{j=0}^{2^{Nn}-1} \sqrt{f^{\mathrm{ind}}(\mathbf{x}_j)\hat{H}(\mathbf{x}_j)} \, |x_1\rangle^n \dots |x_N\rangle^n |1\rangle \\
    &+ \sum_{j=0}^{2^{Nn}-1} \sqrt{(1 - f^{\mathrm{ind}}(\mathbf{x}_j)\hat{H}(\mathbf{x}_j)} \, |x_1\rangle^n \dots |x_N\rangle^n |0\rangle.
\end{aligned}
\end{equation*}
As before, let us denote by $\mathcal{U}_{V^{\mathrm{ind}}} :=  W_{\hat{H}}(\mathcal{A}^{\mathrm{ind}}_{\mathbf{X}} \otimes \mathcal{I})$ to fit the previous quantum state construction into the QAMC description from Section \ref{sec:QAMC}. Note that this formulation allows to use, when available, the exact real density functions to define the independent joint density $f^{\mathrm{ind}}$.

The following theorem is on the accuracy and complexity of the pricing algorithm following the independent approach.

\begin{theorem}[Quantum complexity estimating the option price from independent marginals]
Let $\varrho_c, \epsilon_c \in (0,1)$. Let $c : [0,1]^N \to \R$ be the
density of a copula and suppose that there exists $c'_{\max} \in \mathbb{R}$ such that, for any $i = 1, \ldots, N$ and any $u \in [0,1]^N$,
\begin{equation} \label{eq:BoundDerivativeofCopulaDerivative}
\left| 
\frac{\partial}{\partial u_i}
c(u_1, \ldots, u_N)
\right|
\le c'_{\max}.
\end{equation}
Let $X_1, \ldots, X_N$ be real-valued random variables and suppose
that all the assumptions in Theorem \ref{th:estimation_Fi} are satisfied for every $X_i$. Suppose that we have access to the rotation oracle $W_{\hat{H}}$ for any function in the form of \eqref{PayoffCopulaAdjusted} with $h : \R^N \to [0,1]$. Then, with probability at least $1-\varrho_c$, a QAMC algorithm outputs an $\epsilon_c$-approximation of $\mathbb{E}[h(\mathbf{X})]$, querying $W_{\gamma_k}$

\begin{equation} \label{eq:GammaQueries}
O\left(
\frac{N^2 c'_{\max} \sqrt{I_{\max}} \mathcal{K}_{\max}^2}{\epsilon_c}
\log\!\left( \frac{N\mathcal{K}_{\max}}{\varrho_c} \right)
\right)    
\end{equation}
times, $W_{\hat{H}}$

\begin{equation} \label{eq:PhiQueries}
O\!\left(
\frac{c_{\max}}{\epsilon_c}
\log\!\left( \frac{1}{\varrho_c} \right)
\right)
\end{equation}
times, and $\mathcal{A}_{X_i}$
\begin{equation} \label{eq:AiQueries}
O\!\left(
\frac{N^2 c'_{\max} I_{\max} \mathcal{K}_{\max}^2}{\epsilon_c}
\log\!\left( \frac{N\mathcal{K}_{\max}}{\varrho_c} \right)
+
\frac{N c_{\max}}{\epsilon_c}
\log\!\left( \frac{1}{\varrho_c} \right)
\right)
\end{equation}
times, where  $I_{\max} := \max_{i=1,\ldots,N} (b_i-a_i)$
and $\mathcal{K}_{\max} := \max_{i=1,\ldots,N} \mathcal{K}_i$.
\end{theorem}

\begin{proof}

By applying Theorem \ref{th:estimation_Fi} to each $X_i$ with $\epsilon_i = \frac{\epsilon_c}{2N c'_{\max}}$ and $\varrho_i = \frac{\varrho_c}{2N}$, each estimator $\hat F_i$ satisfies
\begin{equation}\label{eq:FiError}
\forall x\in\mathbb{R},\qquad
\bigl|\hat F_i(x)-F_i(x)\bigr|\le \frac{\epsilon_c}{2N c'_{\max}}
\end{equation}
with probability at least $1-\frac{\varrho_c}{2N}$. For any $\mathbf{x} \in \R^N$, using Taylor’s theorem with \eqref{eq:BoundDerivativeofCopulaDerivative}, we obtain
\begin{align*}
\left|\hat{H}(\mathbf{x}) - H(\mathbf{x})\right|
&= \frac{h(\mathbf{x})}{c_{\max}}\,
\left|\,c\!\left(\hat F_1(x_1),\dots,\hat F_N(x_N)\right)
      -c\!\left(F_1(x_1),\dots,F_N(x_N)\right)\right|
\\
&\le \frac{c'_{\max} h(x)}{c_{\max}}
     \sum_{i=1}^N \left|\hat{F}_i(x_i)-F_i(x_i)\right|
\\
&\le \frac{\epsilon_c}{2c_{\max}}.
\end{align*}
We then have
\begin{equation}\label{eq:errorbetweenPhihatandPhi}
\begin{aligned}
\left|
\mathbb{E}^{\mathrm{ind}}[\hat{H}(\mathbf{X})]
-
\mathbb{E}^{\mathrm{ind}}[H(\mathbf{X})]
\right|
&=
\left|
\sum_{j = 0}^{2^{Nn} - 1}
\left(
\hat{H}(\mathbf{x}_{j}) - H(\mathbf{x}_{j})
\right)
f^{\mathrm{ind}}(\mathbf{x}_{j})
\right|
\\
&\le
\sum_{j = 0}^{2^{Nn} - 1}
\left|
\hat{H}(\mathbf{x}_{j}) - H(\mathbf{x}_{j})
\right|
f^{\mathrm{ind}}(\mathbf{x}_{j})
\\
&\le
\frac{\epsilon_c}{2 c_{\max}},
\end{aligned}
\end{equation}
where, again, $Nn$ denotes the number of qubits employed for the QAMC estimation as described in Section \ref{sec:QAMC} and, recalling that $f^{\mathrm{ind}}(\mathbf{x}) = \prod_{i=1}^N f_i(x_i)$.

On the other hand, the QAE in Theorem \ref{th:qmci} with parameters $\epsilon = \frac{\epsilon_c}{2c_{\max}}$ and $\varrho = \frac{\varrho_c}{2}$ outputs 
$\widehat{\mathbb{E}}^{\mathrm{ind}}[\hat{H}(\mathbf{X})]$ such that
\begin{equation}\label{eq:qmcierrorforPhihat}
\left|
\widehat{\mathbb{E}}^{\mathrm{ind}}[\hat{H}(\mathbf{X})]
-
\mathbb{E}^{\mathrm{ind}}[\hat{H}(\mathbf{X})]
\right|
\le
\frac{\epsilon_c}{2c_{\max}}
\end{equation}
with probability at least \(1 - \frac{\varrho_c}{2}\).

Combining \eqref{eq:errorbetweenPhihatandPhi} and \eqref{eq:qmcierrorforPhihat}, we get
\begin{equation*}
\begin{aligned}
\left| c_{\max} \, \hat{\E}^{\mathrm{ind}}[\hat{H}(\mathbf{X})] - \E_{X}[h(\mathbf{X})] \right|
&\le 
c_{\max} \left(
    \left| \hat{\E}^{\mathrm{ind}}[\hat{H}(\mathbf{X})]
        - \E^{\mathrm{ind}}[\hat{H}(\mathbf{X})] \right|
    +
    \left| \E^{\mathrm{ind}}[\hat{H}(\mathbf{X})]
        - \E^{\mathrm{ind}}[H(\mathbf{X})] \right|
\right) \\
&\le \epsilon_c.
\end{aligned}
\end{equation*}

This holds if every $\hat{F}_i$ satisfies \eqref{eq:FiError} and \eqref{eq:qmcierrorforPhihat}, whose
probability is at least
\begin{equation*}
\left( 1 - \frac{\varrho_c}{2N} \right)^{N}
\left( 1 - \frac{\varrho_c}{2} \right)
\;\ge\; 1 - \varrho_c,
\end{equation*}
by using successively the inequality $(1 - x)^p \ge 1 - px$ for $x \in [0,1]$ and integer $p \in \N$.

Lastly, let us evaluate the query complexity of the algorithm. In estimating each $\hat{F}_i$, $\mathcal{A}_{X_i}$ and $\{ W_{\gamma_k} \}_k$ are queried
\[
O\!\left(
\frac{N c'_{\max} \sqrt{b_i-a_i} \mathcal{K}_i^{2}}{\epsilon_c}
\log\!\left( \frac{N \mathcal{K}_i}{\varrho_c} \right)
\right)
\]
times, and, in estimating all of $\hat{F}_1, \ldots, \hat{F}_d$, this is multiplied by $N$. In the estimate \eqref{eq:qmcierrorforPhihat}, an oracle $\mathcal{A}^{\mathrm{ind}}_{\mathbf{X}}$ loading $f^{\mathrm{ind}}$ and $W_{\hat{H}}$ are called the numbers of times of order \eqref{eq:PhiQueries}. In $\mathcal{A}^{\mathrm{ind}}_{\mathbf{X}}$, the oracles $\mathcal{A}_{X_1}, \ldots, \mathcal{A}_{X_N}$ are called once each, and $N$ times in total. Combining these observations, we reach the query number bounds in \eqref{eq:GammaQueries}, \eqref{eq:PhiQueries}, and \eqref{eq:AiQueries}.

\end{proof}

\subsection{Experimental results}\label{sec:quantum_experimental}

In this section, the performance of the proposed quantum-based methodology is experimentally tested, specifically when applied to estimate the expectations for, on the one hand, the coefficients of the cosine series expansions of the marginal densities (see \eqref{eq:cosine_coef}), and, on the other hand, the final multidimensional option price (see \eqref{eq:multi_option_price} or \eqref{eq:PricingwithCopula}). We will compare the precision convergence patterns of both the CMC estimator given by \eqref{eq:MonteCarlo} and the analogous QAMC, resulting after applying QAE algorithms to the states of the form \eqref{eq:U}. To that end, given a prescribed accuracy, the number of samples $L$ for CMC and the number of queries to the quantum oracle $\mathcal{U}_y,\; y\in\{a_k, V, V^{\mathrm{ind}}\}$ for QAMC are reported. Moreover, since these techniques intrinsically present a random nature, each estimation experiment is repeated $2^5$ times, such that we can then provide statistics like the averages or confidence intervals.

As marginals, we consider the NIG distributions fitted to the market quotes of Credit Agricole, Axa and Michelin (see Section \ref{sec:calibration_results}). The employed calibrated parameters are reported in Figures \ref{fig:CA_Calibrated_IV_PDF}, \ref{fig:Axa_Calibrated_IV_PDF}, and \ref{fig:Michelin_Calibrated_IV_PDF}, for Credit Agricole, Axa, and Michelin, respectively. The remaining market data has been extracted from Euronext as explained in Section \ref{sec:market_data}.

All the experiments have been conducted in a system with processor Intel Core Ultra 9 285H and RAM of 64 GB. The codes are implemented in Python 3.10, and employing the NEASQC: Financial Applications Library \cite{NEASQC_library} under the quantum package myQLM 1.12.2. The quantum simulator relies on C-based linear algebra libraries, becoming an ideal\footnote{It does not include system noise, qubit coherence times, etc.} simulator. As QAE routines, we employ the modified versions of the Real Quantum Amplitude Estimation proposed in \cite{rqae,mrqae} to compute the cosine series coefficients (where the sign of the quantity is relevant) and the Iterative Quantum Amplitude Estimation from \cite{Grinko_2021,Fukuzawa_2023} for the estimation of the option price (assumed positive). Both quantum algorithms come along with rigorous theoretical analysis in terms of error convergence, strictly complying with the order stated in Theorem \ref{th:qmci}.


\subsubsection{Convergence in estimating the density cosine expansion coefficients}

\begin{figure}[h!]
    \centering
   \subfloat[Error in the $k$-th coefficient.]{\includegraphics[width=0.45\linewidth]{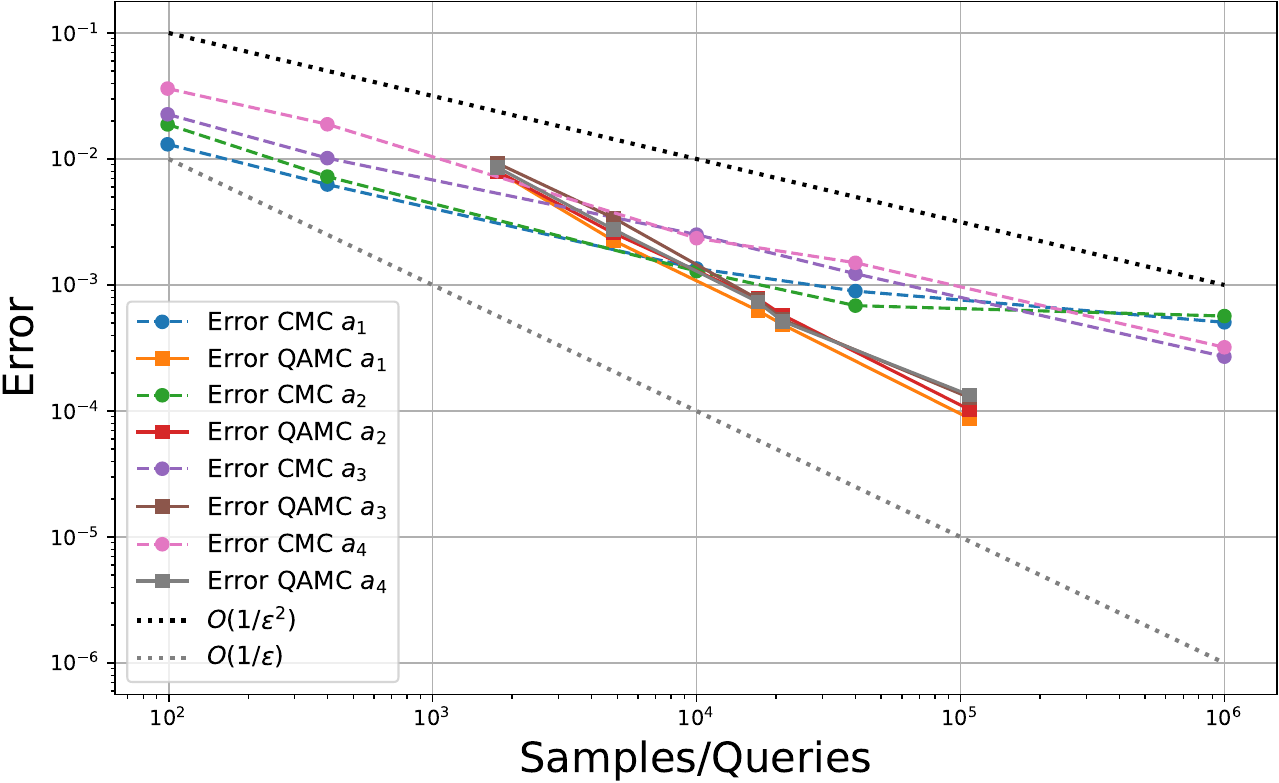}}
    \hspace{1cm}
    \subfloat[Average error of all coefficients.]{\includegraphics[width=0.45\linewidth]{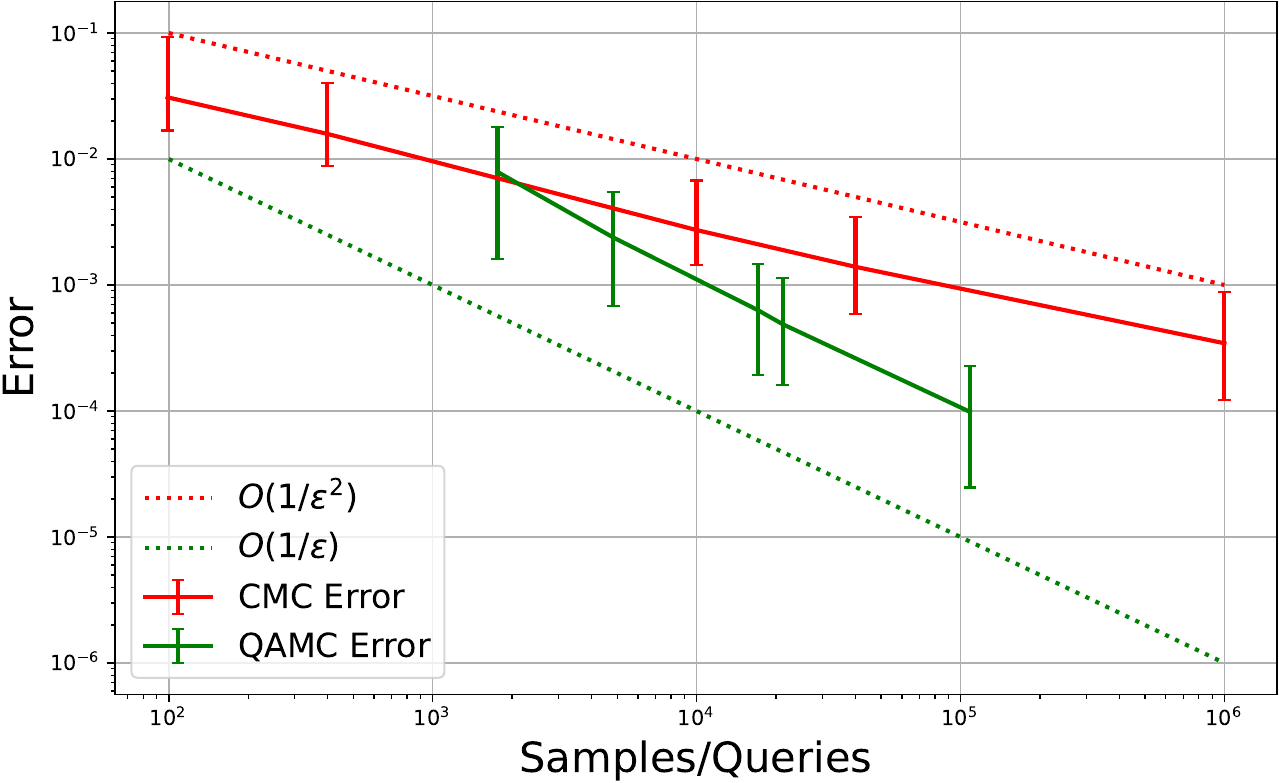}}
    \caption{Convergence in accuracy estimating $a_k$ by CMC and QAMC.}
    \label{fig:convergence_ak}
\end{figure}

In the first experiment, the precision convergence in number of samples and oracle queries for the CMC and QAMC estimators, respectively, is analysed, when applied to recover the calibrated NIG density\footnote{We consider the NIG component of the exponential NIG model, so the densities depicted here correspond to expression \eqref{eq:NIG_density}.} corresponding to AXA. For that, the error between the approximated coefficients and a reference value (given by the Riemann quadrature in \eqref{eq:riemann} computed classically) is considered. The number of discrete points in the quadrature is set to $J=2^5$, which corresponds to $n=5$ qubits employed for the QAMC method. In Figure \ref{fig:convergence_ak}, the obtained results are shown, taking $\mathcal{K}=2^4$. In the left panel, the error convergence lines for the first four (out of sixteen) relevant coefficients (excluding the coefficient $a_0$ which presents exact solution) are depicted as the average of experiment trials. In the right panel, the average error for all the coefficients (jointly with the $90\%$ confidence intervals over the repetitions) are represented. We observe that the CMC estimator deteriorates for higher index coefficients (those with smaller magnitude), while QAMC does not suffer from this issue, due to the natural intrinsic normalization of the amplitudes in a quantum state. All in all, the global expected behaviour is achieved, with QAMC providing a consistent quadratic advantage with respect to CMC.

\begin{figure}[h!]
    \centering
    \subfloat[NIG density with $\mathcal{K}= 2^3$.]{\includegraphics[width=0.32\linewidth]{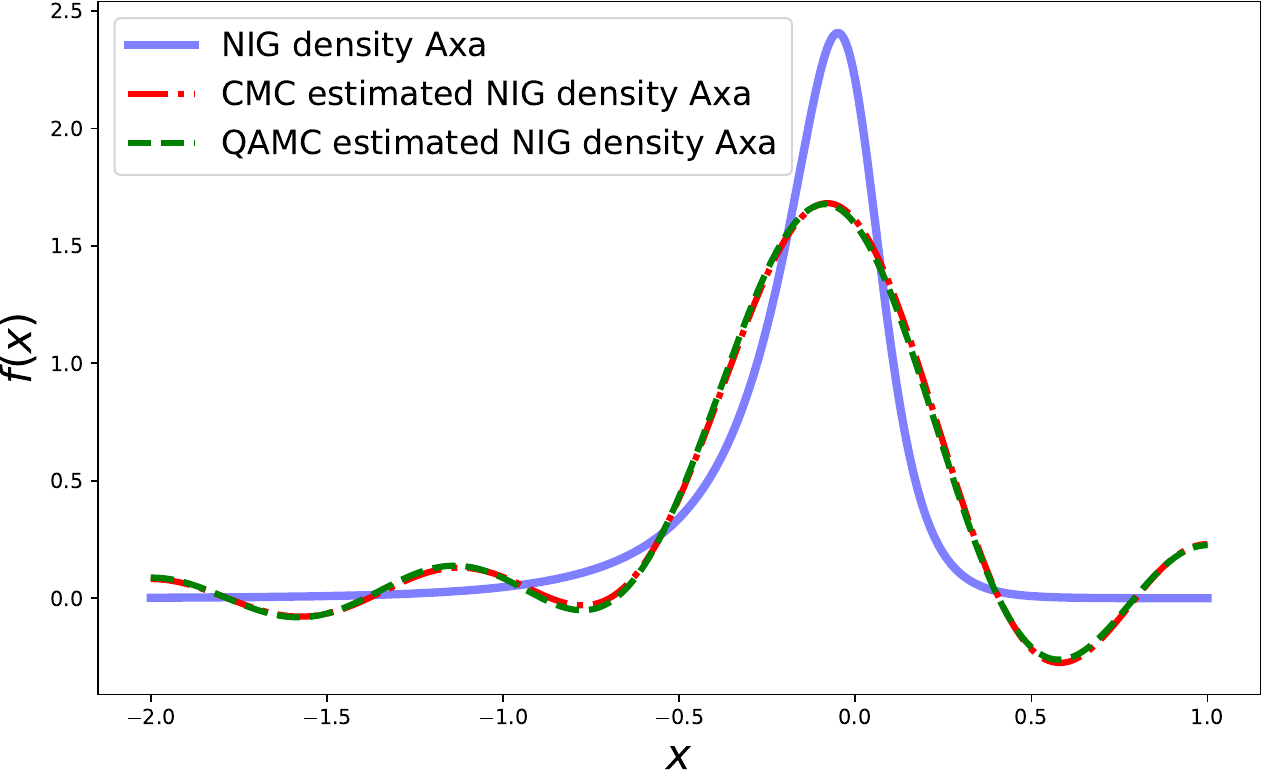}}
    \hspace{0.2cm}
    \subfloat[NIG density with $\mathcal{K} = 2^4$.]{\includegraphics[width=0.32\linewidth]{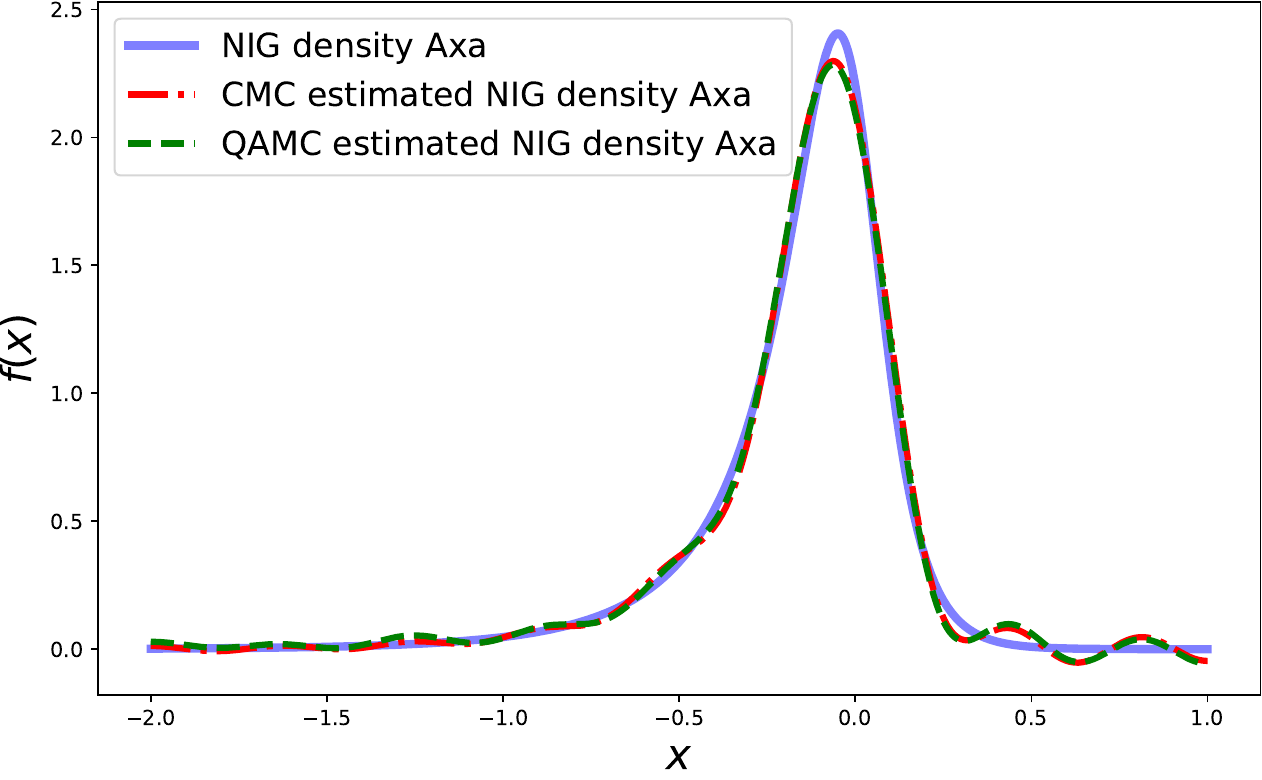}}
    \hspace{0.2cm}
    \subfloat[NIG density with $\mathcal{K} = 2^5$.]{\includegraphics[width=0.32\linewidth]{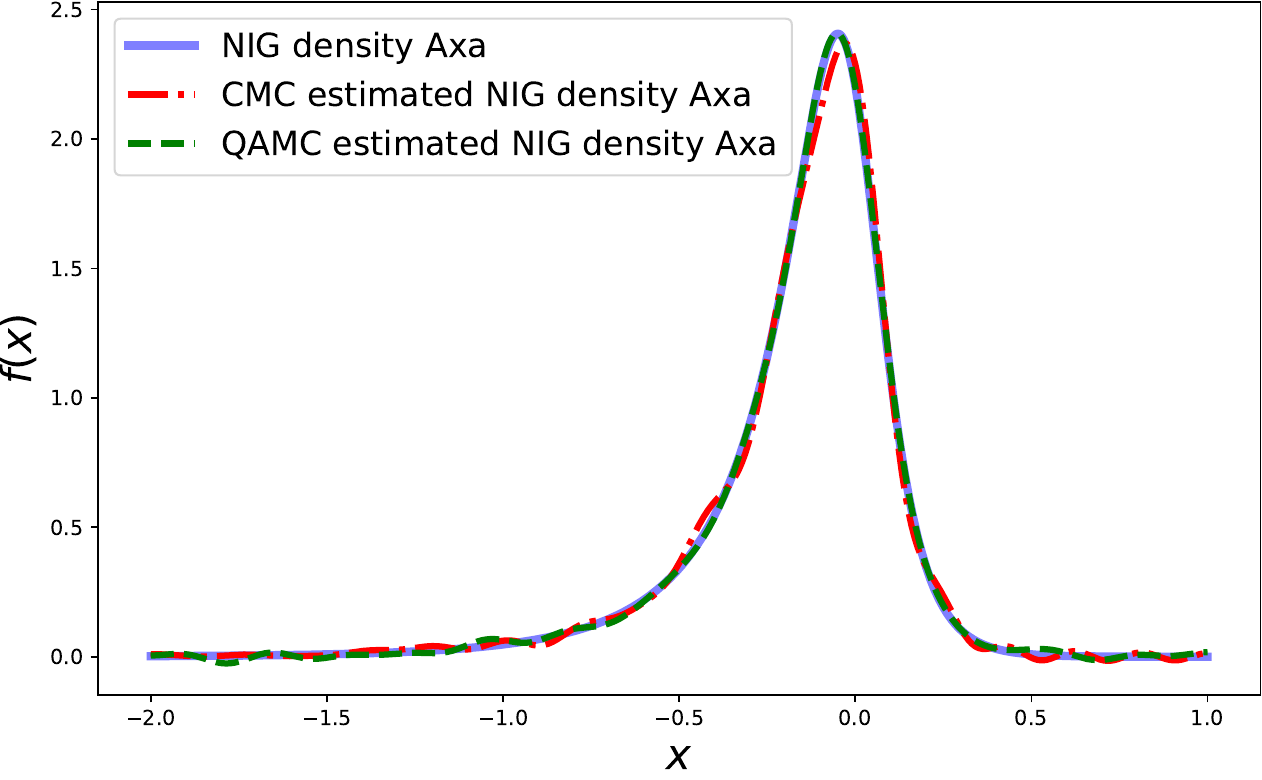}}
    \\
    \subfloat[NIG distribution with $\mathcal{K} = 2^3$.]{\includegraphics[width=0.32\linewidth]{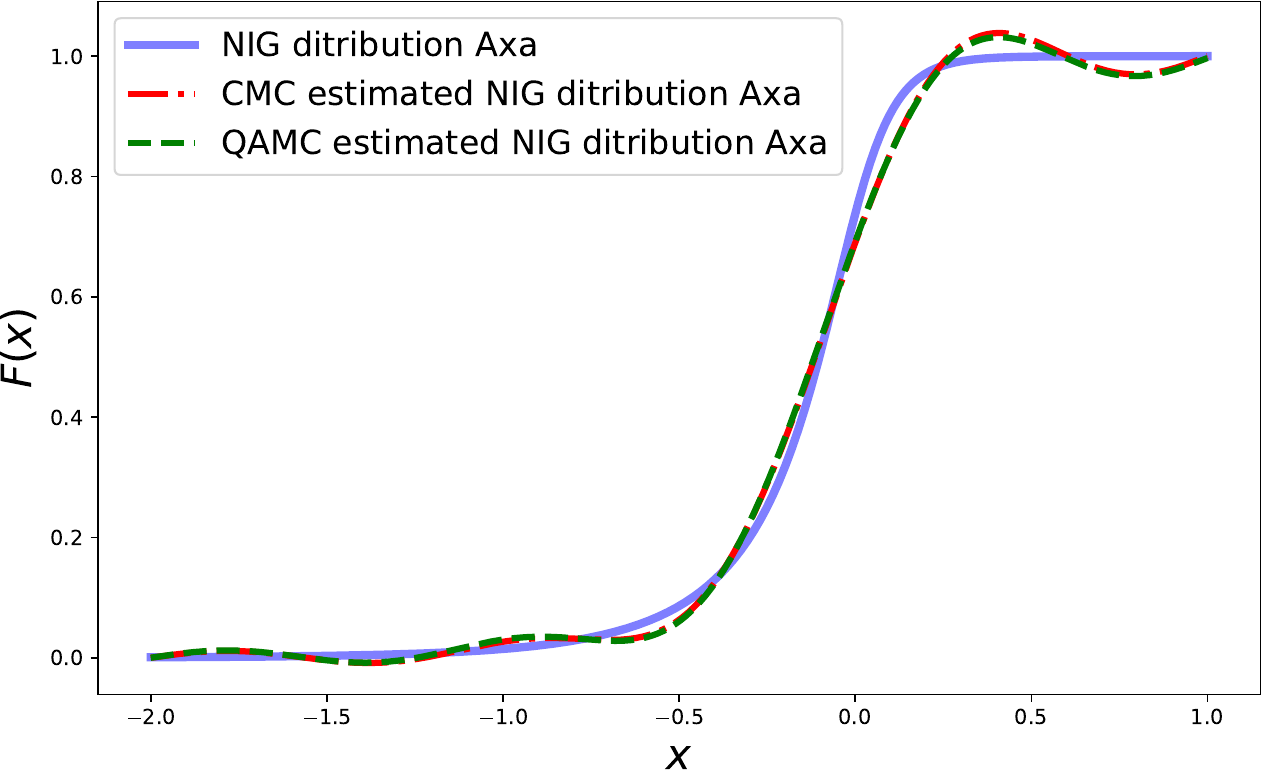}}
    \hspace{0.2cm}
    \subfloat[NIG distribution with $\mathcal{K} = 2^4$.]{\includegraphics[width=0.32\linewidth]{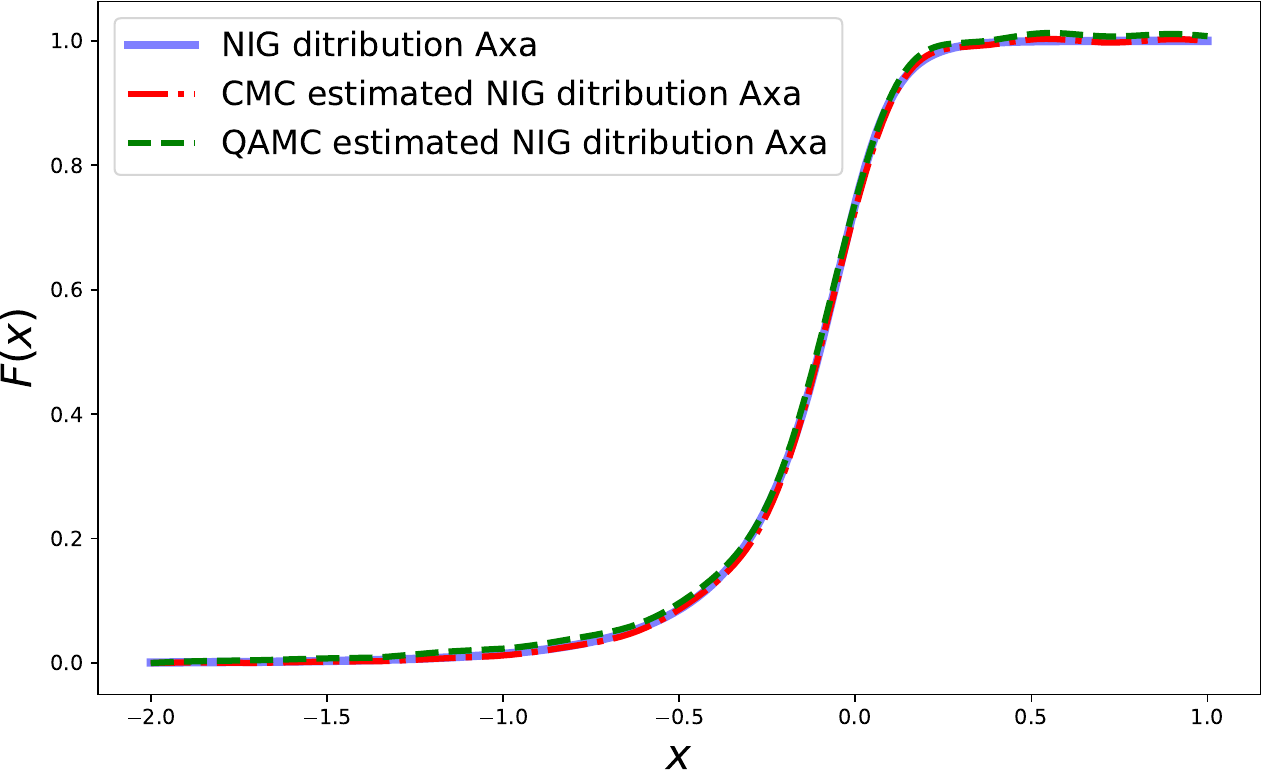}}
    \hspace{0.2cm}
    \subfloat[NIG distribution with $\mathcal{K} = 2^5$.]{\includegraphics[width=0.32\linewidth]{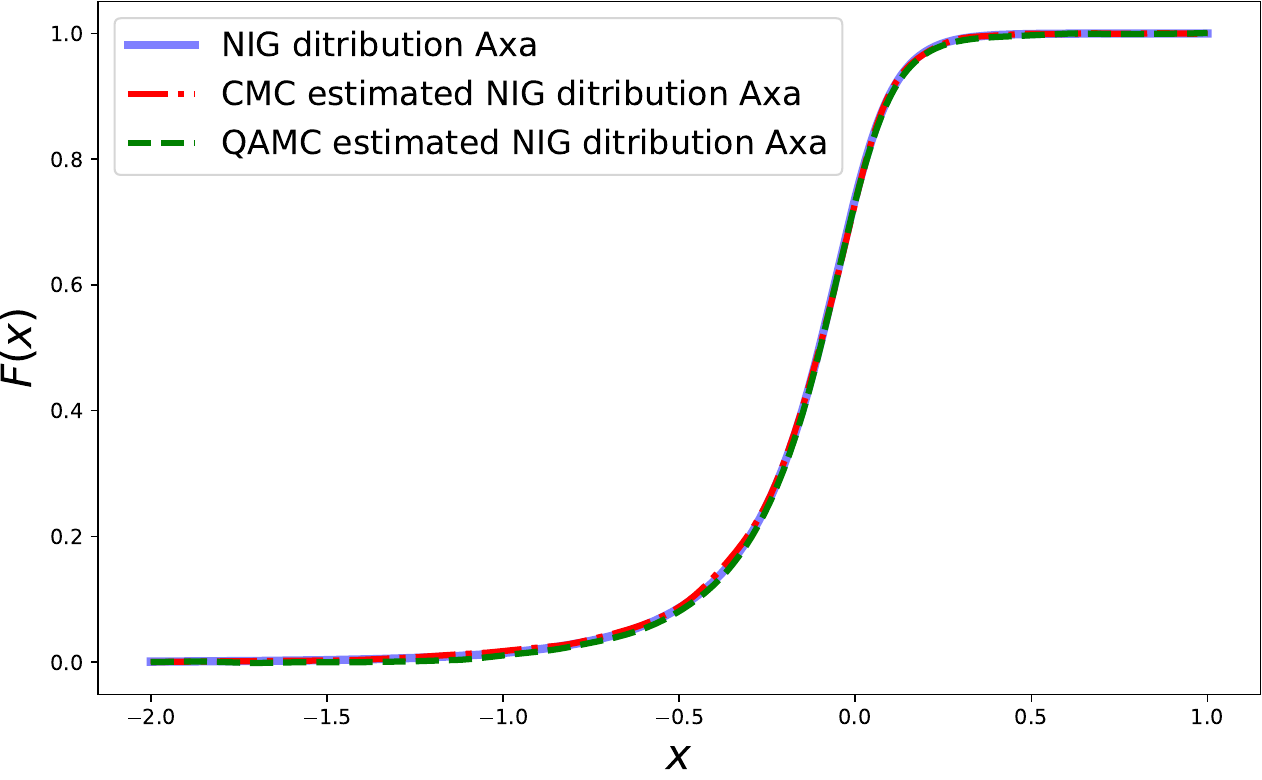}}
    \caption{NIG density and distribution functions for AXA, estimated by CMC and QAMC varying $\mathcal{K}$.}
    \label{fig:density_distribution_K}
\end{figure}

Further, Figure \ref{fig:density_distribution_K} shows the recovered density and distribution functions for an increasing number of expansion terms $\mathcal{K}$ whose coefficients are computed by using the CMC and the QAMC methods with the same number of samples/queries ($\sim 5000$). As we can see, both methods perform very similarly, practically indistinguishably, for lower $\mathcal{K}$s, while, in the case of larger number of terms, the QAMC-based estimations outperform those given by the CMC equivalent. Note as well that, as expected, when $\mathcal{K}$ increases, the global estimations improve.

\subsubsection{Convergence in estimating the option price}

Next, we assess the performance of the QAMC estimator in solving the final multidimensional option valuation problem by comparing it, as in the previous experiment, against the CMC estimator in terms of error convergence in samples/queries. We consider two pricing problems (see Section \ref{sec:multidim_option_pricing_copulas}):
\begin{enumerate}
    \item Spread 1-year expiry call option with AXA and Michelin as underlying assets. The strike is set to and $K=0$ and we model the joint distribution by a Gaussian copula with correlation matrix
    \begin{equation*}
        \Sigma = 
        \begin{pmatrix}
            1 & -0.25 \\
            -0.25 & 1
        \end{pmatrix}.
    \end{equation*}

    \item Arithmetic basket 1-year expiry call option with AXA, Credit Agricole and Michelin as underlying assets. The strike is set to $K=25$ and we model the joint distribution by a Gaussian copula with correlation matrix
    \begin{equation*}
        \Sigma = 
        \begin{pmatrix}
            1 & -0.2 & -0.25 \\
            -0.2 & 1 & -0.15 \\
            -0.25 & -0.15 & 1 
        \end{pmatrix}.
    \end{equation*}
\end{enumerate}

\begin{figure}[h!]
    \centering
    \subfloat[Spread call option]{\includegraphics[width=0.48\linewidth]{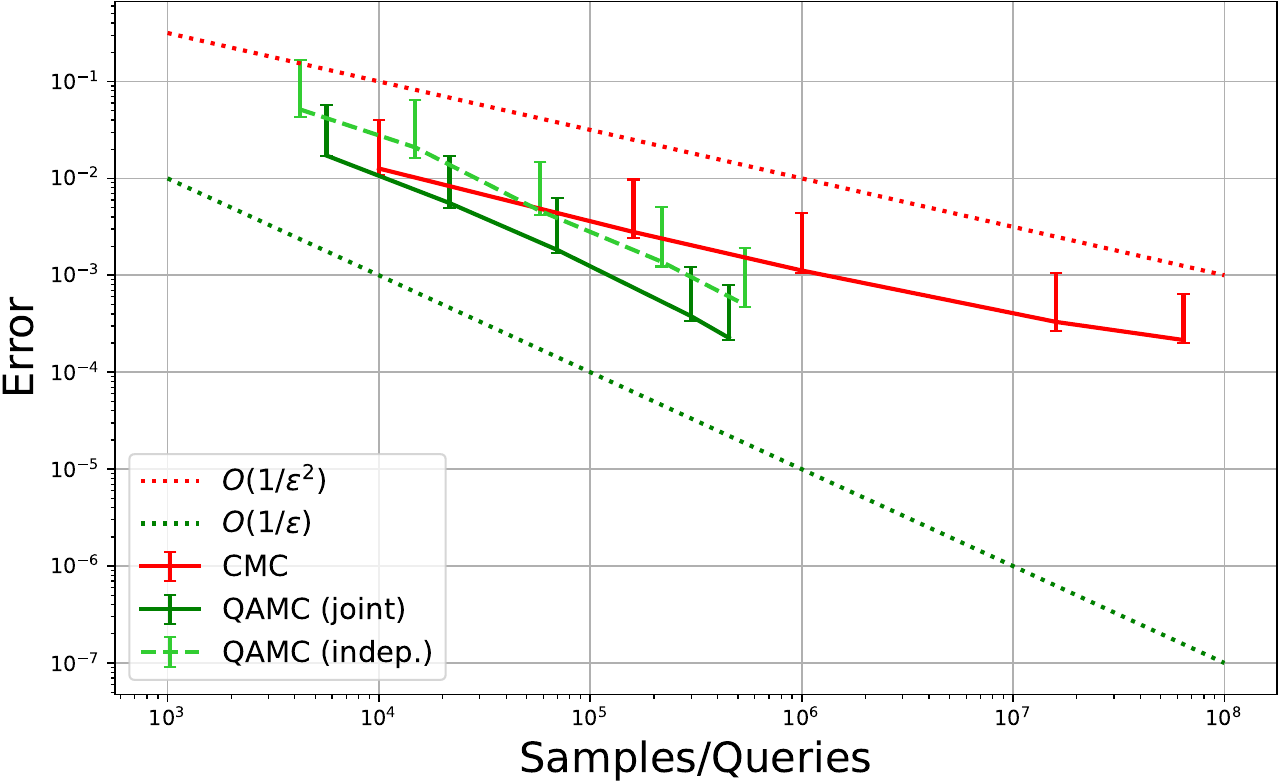}}
    \hspace{0.5cm}
    \subfloat[Arithmetic basket call option]{\includegraphics[width=0.48\linewidth]{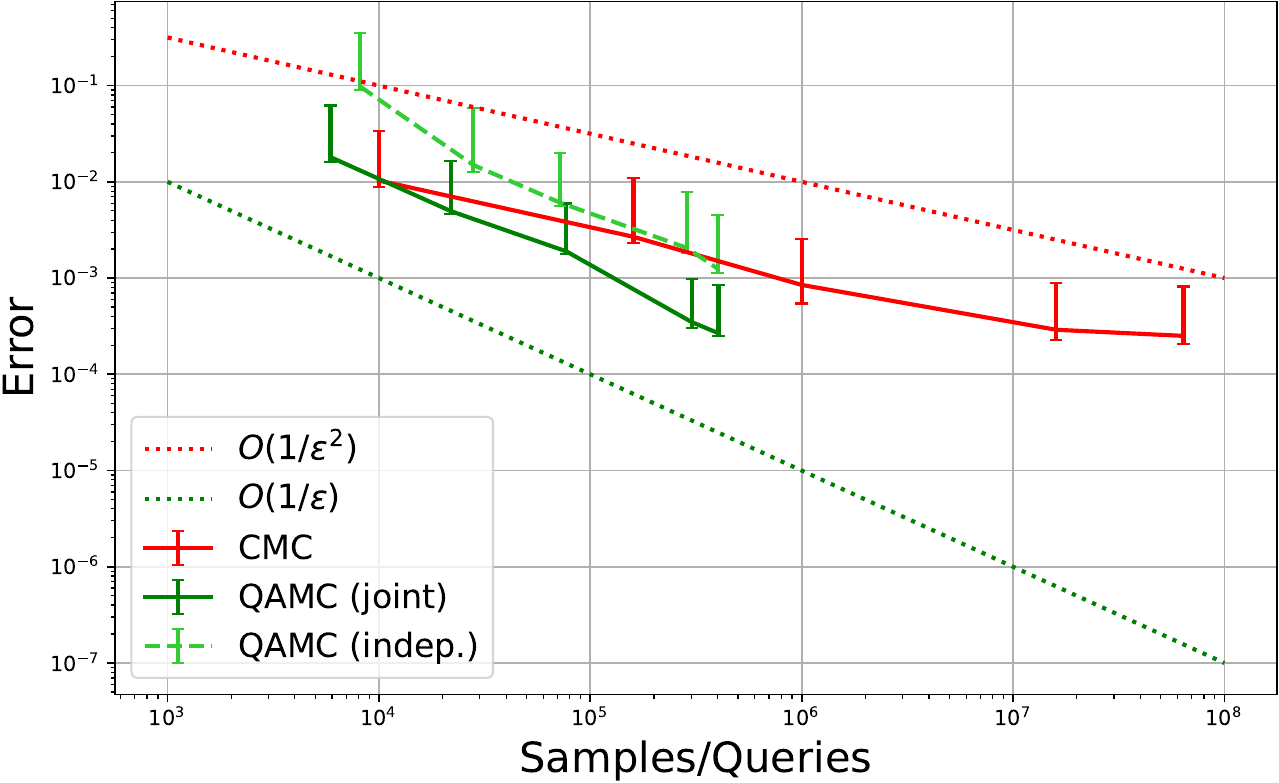}}
    \caption{Convergence in accuracy estimating the multidimensional option price.}
    \label{fig:convergence_price}
\end{figure}

Again, the accuracy in the estimation is measured against a reference price obtained via a classically computed Riemann sum. Due to the extremely high computational demand of the considered quantum simulator, we adapt the number of employed discrete points to the dimensionality of the problem at hand. Then, for the spread option, we choose $J = 2^3$ points in each space direction ($N=2$), while, in the case of the arithmetic basket option valuation, we select $J = 2^2$ discrete points per dimension ($N=3$). This then entails that, in both cases, we employ a total of $Nn=6$ qubits to apply the QAMC technique. In order to isolate the error due to the computation of the final price, the marginal densities are recovered with $\mathcal{K}=2^7$ cosine series coefficients, computed classically (so no quantum-related approximation error arises from them). In Figure \ref{fig:convergence_price}, the accuracy convergence results for CMC and QAMC estimators (utilising both joint and independent formulations) are presented for spread call and arithmetic basket call options in the left and right panels, respectively. Again, we show the average estimation among several repetitions, as well as the $90\%$ confidence interval.

We can extract the following insights from the pricing experiments:
\begin{itemize}
    \item Both CMC and QAMC algorithms converge at their theoretical orders, namely, $1/\epsilon^2$ and $1/\epsilon$, respectively, which, again, empirically demonstrates the quadratic improvement provided by the QAMC-based solutions as alternatives to the CMC versions in multidimensional option pricing.

    \item Although keeping the order of convergence, for this specific valuation problem, the QAMC relying on the joint formulation significantly outperforms the independent analogous, showing a lower intercept in terms of the number of queries.

    \item The CMC convergence even deteriorates for larger number of samples, suggesting that it might be saturating.

    \item In practical terms, when high accuracy is required (below $10^{-3}$), QAMC needs $10-100$ fewer samples/queries than CMC, to achieve a prescribed precision.
\end{itemize}

\section{Conclusions}\label{sec:conclusions}

This work presents a comprehensive framework for multi-asset option pricing that integrates market-consistent modelling with quantum-accelerated computation. By calibrating NIG marginals to real option quotes and coupling them through a Gaussian copula, we construct arbitrage-free joint distributions capable of capturing skewness and fat tails observed in equity markets. The proposed calibration procedure, supported by theoretical guarantees of existence and continuity, achieves high accuracy with minimal pricing errors across multiple assets.

On the computational front, we demonstrate that QAMC methods, based on QAE, deliver the expected quadratic improvement in convergence compared to CMC. Empirical experiments confirm that QAMC requires significantly fewer queries (by one to two orders of magnitude) for comparable precision, particularly in high-dimensional settings. These results validate the practical feasibility of quantum algorithms for complex derivative pricing and highlight their potential to overcome scalability limitations inherent in classical approaches.

Beyond immediate performance gains, this work underscores the importance of combining arbitrage-aware modelling with quantum techniques to ensure both financial soundness and computational efficiency. Future research should explore richer dependence structures beyond Gaussian copulas, extend the pipeline to path-dependent payoffs, and investigate hardware implementations to assess real-world resource constraints. By bridging rigorous market modelling and quantum computing, this study contributes a foundational step toward deployable quantum solutions in quantitative finance.

\section*{Acknowledgements}

Both authors thank the Euronext data support team for their kind assistance in providing and clarifying the option data used in our analysis.

Á. Leitao acknowledges the funding from the Ministry of Science and Innovation of Spain through the Ramón y Cajal 2022 grant and the program with reference PID2022-141058OB-I00, and from the Department of Education, Science, Universities, and Vocational Training of the Xunta de Galicia through the programs with references ED451C 2022/047 and ED431F 2025/032, as well as the support from CITIC, as a centre accredited for excellence within the Galician University System and a member of the CIGUS Network, receiving subsidies from the Department of Education, Science, Universities, and Vocational Training of the Xunta de Galicia. Additionally, it is co-financed by the EU through the FEDER Galicia 2021-27 operational program (ref. ED451G 2023/01).

\bibliographystyle{abbrv}
\bibliography{references}

\end{document}